\documentclass[a4paper,11pt]{article}
\pdfoutput=1 % if your are submitting a pdflatex (i.e. if you have
\usepackage[T1]{fontenc} % if needed
\usepackage{etoolbox}
\usepackage{amssymb}
\usepackage{amsmath}
\usepackage{tcolorbox}
\usepackage{graphics}
\usepackage[final]{pdfpages}
\usepackage{slashed}
\usepackage{amssymb}
\usepackage{epstopdf}
\catcode`@11
\def\seceqaa{\@addtoreset{equation}{section}
\def\theequation{A\arabic{equation}}}
\def\seceqbb{\@addtoreset{equation}{section}
\def\theequation{B\arabic{equation}}}
\catcode`@11

\title{}

\date{\today}

\begin{document}

\begin{titlepage}
%\title{}
\begin{center}
%{\large \bf
{\large \bf The QCD Trace Anomaly at Strong Coupling from M-Theory}\\
Aalok Misra$^{(a)}$\footnote{e-mail: aalok.misra@ph.iitr.ac.in}
and Charles Gale$^{(b)}$\footnote{email:gale@physics.mcgill.ca }\\
(a) Department of Physics, Indian Institute of Technology,
Roorkee - 247 667, Uttarakhand, India\\
(b) Department of Physics, McGill University, 3600 University St, Montr\'eal, QC H3A 2T8, Canada

\end{center}
\thispagestyle{empty}

\begin{abstract} 
Obtaining a lattice-consistent result for  the temperature dependence of the QCD conformal anomaly  from a top-down M-theory dual (valid) for all temperatures - {\it both}, $T<T_c$ and $T>T_c$ - of thermal QCD at intermediate gauge coupling, has been missing in the literature. We fill this gap by addressing this issue  from  the  M-theory uplift of the SYZ type IIA mirror  at intermediate gauge/string coupling ( both obtained in \cite{MQGP})  of the UV-complete type IIB holographic dual of large-$N$ thermal QCD of \cite{metrics}, and comparing with the very recent lattice results of \cite{Petreczky-et-al}. Estimates of the ${\cal O}(R^4)$ higher derivative corrections in the $D=11$ supergravity action relevant to considering the aforementioned M theory uplift in the intermediate 't Hooft coupling (in addition to gauge coupling)  limit, are also presented.  We also show that after a tuning of the (small) Ouyang embedding parameter and radius of a blown-up $S^2$ when expressed in terms of the horizon radius, a QCD deconfinement temperature $T_c=150$~MeV from a Hawking-Page phase transition at vanishing baryon chemical potential consistent with lattice QCD in the heavy-quark limit, can be obtained.
%We believe, a top-down holographic computation of the QCD trace anomaly, was previously missing in the literature.}
\end{abstract}
\end{titlepage}

\section{Introduction}

One of the breakthroughs of the relativistic heavy ion program has been the realization that the production of hadronic matter in extreme conditions of temperature and density -- as created during the high-energy collision of large nuclei -- can be well modelled and understood using numerical simulations relying on relativistic fluid dynamics \cite{Gale:2013da}. In that context, the QCD equation of state (EOS) is an  essential quantity. Nonperturbative calculations based on lattice QCD have now confirmed the fact that at vanishing baryonic density the transition between partonic degrees of freedom and those in the confined sector is in fact a rapid crossover \cite{Borsanyi:2013bia,Bazavov:2014pvz,Petreczky-et-al} occurring in the vicinity of $T \simeq 150$ MeV. For higher baryon densities and slightly lower temperatures, lattice calculations have proven to be challenging because of the notorious sign problem \cite{KG}. Some progress has nevertheless been made, through a variety of techniques \cite{KG}. At still lower temperatures and high densities,  investigations of the hadronic EOS through several different approaches suggest a first-order chiral symmetry restoration phase transition \cite{Halasz:1998qr,Berges:1998rc,Schafer:1999hp}. These results and others like them have fuelled much of the interest in the search for a critical end point (CEP) and the initiation of a beam energy scan (BES) at RHIC \cite{Keane:2017kdq}. Thus, in parallel with theoretical work, experimental explorations can be used to uncover subtle but fundamental features of the EOS, such as the existence of a possible critical point and of genuine thermodynamic phase transitions \cite{Caines:2017vvs}. Finally, the importance of the hadronic EOS is not restricted to the field of relativistic heavy-ion collisions. The EOS is responsible for the bulk properties of dense stellar objects such as neutron stars. It also affects their cooling properties, which probes the particle content and the state of the matter present in their core \cite{Cumming:2016weq,Brown:2017gxd}. 

The  EOS is an integral part of the energy-momentum tensor, $T^{\mu \nu}$. In a classical theory without any dimentionful parameters, a scale transformation leaves the action invariant, and conversely leads to a traceless energy momentum tensor: $T^\mu_\mu = 0$. This is the case for classical, massless, Yang-Mills theory. However, quantum effects will spoil the conservation of the dilatation current, and make the theory scale-dependent \cite{Peskin:1995ev}, as clearly shown by the running of the coupling, $g$, via the $\beta$-function: $\beta(g) = \mu\, \partial g  /\partial \mu$. Then, the Yang-Mills $T^{\mu \nu}$ satisfies 
\begin{eqnarray}
T^\mu_\mu = \frac{\beta(g)}{2 g} F^{\mu \nu\, a}F^a_{\mu \nu}
\end{eqnarray}
where $a$ is a color index. 

This discussion brings us to the core of this paper, and to its two-fold intent. First, it is clear that calculations of the hadronic EOS clearly requires a treatment which goes beyond perturbation theory. In this context, the gauge-gravity duality  \cite{Maldacena:1997re} offers an appealing set of techniques which render strong coupling calculations analytically feasible. The original formulation of the duality was AdS/CFT: the field theory sector was conformal. More recently, extensions into families of theories which break conformal invariance have been actively pursued. We now briefly describe the approach used in this work and the path which lead to its development; details of the former are given in Section {\bf 2}.

Gauge/gravity duality has proved to be very useful in understanding the properties of (thermal) QCD-like theories.
%Building up on the Klebanov-Witten \cite{KW}, Klebanov-Nekrasov \cite{KN}  and  Klebanov-Tseytlin \cite{KT} models,  a logarithmic RG flow just like QCD was obtained in the non-conformal Klebanov-Strassler model \cite{KS} by considering $M$ fractional $D3$ branes along with $N$  $D3$ branes in a conifold geometry  wherein the IR geometry was modified resulting in a deformed conifold.
%In fact the AdS/CFT correspondence mentioned above is valid at zero temperature. At finite temperature the situation is different on the gravity side of the correspondence.
The first top-down (type IIA) holographic dual of large-$N$ QCD though catering only to the IR, was given in \cite{Sakai_Sugimoto}. A UV-complete (type IIB) holographic dual of large-$N$ thermal QCD, was constructed in \cite{metrics}. It is believed that large-$N$ thermal QCD laboratories like strongly coupled QGP (sQGP) require not only a large 't Hooft coupling but also a intermediate gauge coupling \cite{Natsuume}. Holographic models based on this assumption, therefore necessarily require addressing this limit from M theory. It is known that QCD possesses a rapid crossover from a confining phase to a non-confining phase at $T \simeq T_c$,
%At sufficiently high temperature i.e. at $T\gg T_c $, the theory is weakly coupled. 
and to explore the physics of QCD at $T\approx T_c$, we have to take a look at the strongly coupled regime of the theory. The holographic study of large-$N$ thermal QCD at intermediate coupling, was initiated in \cite{MQGP} which presented a M-theory uplift of the SYZ type IIA mirror (in the spirit of \cite{syz,M. Becker et al [2004]}) of a string theoretic dual of large-$N$ thermal QCD-like theories at intermediate gauge/string coupling as part of the `MQGP' limit of \cite{MQGP}. In this limit, the temperature dependence of a  variety of transport coefficients have been calculated in \cite{EPJC-2}. On the holographic phenomenology front, lattice/PDG-compatible glueball and \mbox{(pseudo-)vector} and \mbox{(pseudo-)scalar} meson masses as well as (exotic scalar)glueball-to-meson decay widths were calculated in \cite{Sil+Yadav+Misra-glueball,mesons_0E++-to-mesons-decays}.

The QCD conformal anomaly and its temperature dependence are important quantities to be studied in the context of, e.g., relativistic heavy ion collisions. In this paper, we will describe how to evaluate the same and obtain, in particular, the temperature dependence of the trace anomaly from M-theory and compare our results with recent lattice results. Note, to the best of our knowledge, there is no precedence of studing the QCD conformal/trace anomaly from a top-down M theory dual (inclusive of higher derivative (${\cal O}(R^4)$) corrections corresponding to considering the intermediate 't-Hooft coupling limit) at low {\it and} high temperatures consistent with recent lattice results \footnote{See, e.g. \cite{Nitti_et_al} for earlier attempts at matching trace anomaly in bottom-up holographic models, with (older) lattice results; also see \cite{trace-anomaly-bottum-up-2,Gursoy,Bayona_et_al,Ficnar_et_al}.}. 

The remainder of this paper is organized as follows. Section {\bf 2} is a brief review of (the UV complete) string/M-theory holographic dual of large-$N$ thermal QCD as constructed in \cite{metrics} (type IIB) and \cite{MQGP,NPB}(type IIA and M-theory) to make this paper self contained. Section {\bf 3} discusses obtaining a lattice-compatible $T_c$ from the type IIB holographic dual as constructed in \cite{metrics} from a Hawking-Page phase transition at zero chemical potential (improving upon a similar computation done earlier in \cite{EPJC-2}). Section {\bf 4} has to do with a holographic computation of the QCD trace anomaly from M theory. This  is partitioned into two sub-sections - {\bf 4.1} is on high temperatures, i.e., $T>T_c$ and {\bf 4.2} is on low temperatures, i.e., $T<T_c$. Section {\bf 5}, apart from summarizing the results, discusses a very crucial point as regards compatibility of our results with lattice computations. The Hawking- Page phase transition in our computation in Section {\bf 3} occurs at zero baryon chemical potential $\mu_C$ and one expects a smooth cross-over for a non-zero $\mu_C$ above a critical value of $\mu_C$  which is exactly the opposite of what one generically expects from (lattice) QCD. We argue that our holographic gravity dual computation can still be justified in the heavy quark limit. There are two appendices.  Appendix {\bf A} discusses the evaluation of the baryon chemical potential and the DBI action on the flavor $D7$-branes. Appendix {\bf B} summarizes the definitions relevant to the ${\cal O}(R^4)$ terms in the $D=11$ SUGRA action

\section{String/M-Theory Dual of Thermal QCD - A Review of \cite{metrics,MQGP}}

In this section, we provide a short review of a UV complete type IIB holographic dual ({\it the only one we are aware of}) of large-$N$ thermal QCD constructed in \cite{metrics}, its Strominger-Yau-Zaslow (SYZ) type IIA mirror at intermediate string coupling and its subsequent M-theory uplift constructed in \cite{MQGP, NPB}. 

\begin{enumerate}
\item {\bf UV-complete holographic dual of large-$N$ thermal QCD as constructed in  \cite{metrics}}: The UV-complete holographic dual of large-$N$ thermal QCD as constructed in  \cite{metrics}, subsumed the zero-temperature Klebanov-Witten model \cite{KW}, the non-conformal Klebanov-Tseytlin model \cite{KT}, its IR completion as given in the Klebanov-Strassler model \cite{KS} and Ouyang's \cite{ouyang} inclusion  of flavor in the same, 
%\footnote{See \cite{Leo-i, Leo-ii} for earlier attempts at studying back-reacted $D3/D7$ geometry at zero temperature.}
as well as the non-zero temperature/non-extremal version of \cite{Buchel} (but the non-extremality/black hole function and the ten-dimensional warp factor vanished simultaneously at the horizon radius), \cite{Gubser-et-al-finitetemp} (valid only at large temperatures)  and \cite{Leo-i,Leo-ii} (addressing the IR), in the absence of flavors. The following summarizes the main features of \cite{metrics}.

\begin{itemize}
\item {\bf Brane construct of \cite{metrics}}:
The type IIB string dual of \cite{metrics} consists of $N$ $D3$-branes placed at the tip of six-dimensional conifold, $M\ D5$-branes wrapping the vanishing $S^2$ and $M\ \overline{D5}$-branes  distributed along the resolved $S^2$ placed at antipodal points relative to the $M$ $D5$-branes. Let us denote the average $D5/\overline{D5}$ separation  by ${\cal R}_{D5/\overline{D5}}$.  Roughly, $r>{\cal R}_{D5/\overline{D5}}$, would be the UV.    The $N_f\ D7$-branes, holomorphically embedded via Ouyang embedding \cite{ouyang} in the resolved conifold geometry, ``smeared"/delocalized along the angular directions $\theta_{1,2}$ as mentioned below (\ref{h_small_r}), are present in the UV, the IR-UV interpolating region and dip into the (confining) IR (but do not touch the $D3$-branes with the shortest $D3-D7$ string corresponding to the lightest quark). In addition, $N_f\ \overline{D7}$-branes are present in the UV and the UV-IR interpolating region for the reason given below. The following table summarizes the aforementioned brane construct wherein $S^2(\theta_1,\phi_1)$ denotes the vanishing two-sphere and (NP/SP of) $S^2_a(\theta_2,\phi_2)$ is the (North Pole/South Pole of the) resolved/blown-up two-sphere - $a$ being the radius of the blown-up $S^2$ - and $r_{\rm UV}$ is the UV cut-off and $\frac{\epsilon}{\left({\cal R}_{D5/\overline{D5}} - |\mu_{\rm Ouyang}|^{\frac{2}{3}}\right)}<1$. Also, $\mu_{\rm Ouyang}$ is the Ouyang embedding parameter that is defined in (\ref{Ouyang-definition}) while describing the embedding of the flavor $D7$-branes in the resolved conifold geometry.
\begin{table}[h]
\begin{center}
\begin{tabular}{|c|c|c|}\hline
&&\\
S. No. & Branes & World Volume \\ 
&&\\ \hline
&&\\
1. & $N\ D3$ & $\mathbb{R}^{1,3}(t,x^{1,2,3}) \times \{r=0\}$ \\
&&\\  \hline
&&\\
2. & $M\ D5$ & $\mathbb{R}^{1,3}(t,x^{1,2,3}) \times \{r=0\} \times S^2(\theta_1,\phi_1) \times {\rm NP}_{S^2_a(\theta_2,\phi_2)}$ \\ 
&&\\  \hline
&&\\
3. & $M\ \overline{D5}$ & $\mathbb{R}^{1,3}(t,x^{1,2,3}) \times \{r=0\}  \times S^2(\theta_1,\phi_1) \times {\rm SP}_{S^2_a(\theta_2,\phi_2)}$ \\ 
&&\\  \hline
&&\\
4. & $N_f\ D7$ & $\mathbb{R}^{1,3}(t,x^{1,2,3}) \times \mathbb{R}_+(r\in[|\mu_{\rm Ouyang}|^{\frac{2}{3}},r_{\rm UV}])  \times S^3(\theta_1,\phi_1,\psi) \times {\rm NP}_{S^2_a(\theta_2,\phi_2)}$ \\ 
&&\\  \hline
&&\\
5. & $N_f\ \overline{D7}$ & $\mathbb{R}^{1,3}(t,x^{1,2,3}) \times \mathbb{R}_+(r\in[{\cal R}_{D5/\overline{D5}}-\epsilon,r_{\rm UV}]) \times S^3(\theta_1,\phi_1,\psi) \times {\rm SP}_{S^2_a(\theta_2,\phi_2)}$ \\ 
&&\\  \hline
\end{tabular}
\end{center}
\caption{The Type IIB Brane Construct of \cite{metrics}}
\end{table}

\item
In the UV, one has $SU(N+M)\times SU(N+M)$ color gauge group and $SU(N_f)\times SU(N_f)$ flavor gauge group. There occurs a partial Higgsing of $SU(N+M)\times SU(N+M)$ to $SU(N+M)\times SU(N)$ as one goes from $r>{\cal R}_{D5/\overline{D5}}$  to $r<{\cal R}_{D5/\overline{D5}}$. This happens because in the IR, at low energies, i.e., at energies less than  ${\cal R}_{D5/\overline{D5}}$, the $\overline{D5}$-branes are integrated out resulting in the reduction of the rank of one of the product gauge groups (which is $SU(N + {\rm number\ of}\ D5-{\rm branes})\times SU(N + {\rm number\ of}\ \overline{D5}-{\rm branes})$). By the same token, the $\overline{D5}$-branes are ``integrated in" in the UV, resulting in the conformal Klebanov-Witten-like $SU(M+N)\times SU(M+N)$ product color gauge group \cite{KW}.  

\item
The pair of gauge couplings, $g_{SU(N+M)}$ and $g_{SU(N)}$, were shown in \cite{KS} to flow  oppositely; in fact the flux of the NS-NS B through the vanishing $S^2$ is the obstruction to obtaining conformality which is why $M$ $\overline{D5}$-branes were included in \cite{metrics} to cancel the net $D5$-brane charge in the UV. Further, as the $N_f$ flavor $D7$-branes enter the RG flow of the gauge couplings via the dilaton (see (\ref{dilaton})), their contribution therefore needs to be canceled by $N_f\ \overline{D7}$-branes which is the reason for their inclusion in the UV in \cite{metrics}. The RG flow equations for the gauge coupling $g_{SU(N+M)}$ - corresponding to the gauge group of a relatively higher rank - can be used to show that the same flows towards strong coupling, and the $SU(N)$ gauge coupling flows towards weak coupling. One can show that the strongly coupled $SU(N+M)$ is Seiberg-like dual to weakly coupled  $SU(N-(M - N_f))$ \footnote{The Seiberg duality (cascade) is applicable for supersymmetric theories. For non-supersymmetric theories such as the holographic dual we are working with, the same is effected via a radial rescaling: $r\rightarrow e^{-\frac{2\pi}{3g_sM_{\rm eff}}}r$ \cite{ouyang} under an RG flow from the UV to the IR.}.  

 \item
{\bf Obtaining} ${\bf N_c=3}$: In the IR, at the end of a Seiberg-like duality cascade,  the number of colors $N_c$ gets identified with $M$, which in the `MQGP limit' to be discussed below, can be tuned to equal 3. This is briefly explained now. One can identify $N_c$ with 
\begin{equation}
\label{Neff-def}
N_{\rm eff}(r) + M_{\rm eff}(r),
\end{equation}
 where $N_{\rm eff}(r)$ is defined via 
\begin{equation}
\label{Ftilde5-def}
\tilde{F}_5\equiv dC_4 + B_2\wedge F_3 = {\cal F}_5 + *{\cal F}_5,
\end{equation}
wherein ${\cal F}_5\equiv N_{\rm eff}{\rm Vol}({\rm Base\ of\ Resolved\ Warped\ Deformed\ Conifold})$, and 
\begin{equation}
\label{Meff-def}
M_{\rm eff} = \int_{S^3}\tilde{F}_3
\end{equation}
 (the $S^3$ being dual to $\ e_\psi\wedge\left(\sin\theta_1 d\theta_1\wedge d\phi_1 - B_1\sin\theta_2\wedge d\phi_2\right)$, wherein $B_1$ is an asymmetry factor defined in \cite{metrics}; $e_\psi\equiv d\psi + {\rm cos}~\theta_1~d\phi_1 + {\rm cos}~\theta_2~d\phi_2$) where \cite{M(r)N_f(r)-Dasgupta_et_al}: 
\begin{equation}
\label{F3-def}
\tilde{F}_3 (\equiv F_3 - \tau H_3)\propto M(r)\equiv M\frac{1}{1 + e^{\alpha(r-{\cal R}_{D5/\overline{D5}})}}, \alpha\gg1.
\end{equation}
 The effective number $N_{\rm eff}$ of $D3$-branes varies between $N\gg1$ in the UV and 0 in the deep IR, and the effective number $M_{\rm eff}$ of $D5$-branes varies between 0 in the UV and $M$ in the deep IR. Hence, $N_c$ varies between $M$ in the deep IR and a large value [ in the MQGP limit of (\ref{MQGP_limit}) for a large value of $N$] in the UV.  Hence, at very low energies, the number of colors $N_c$ can be approximated by $M$, which in the MQGP limit is taken to be finite and can hence be taken to be equal to three. Additionally, one can set $N_f=2+1$ for comparison  with \cite{Petreczky-et-al}. Hence, in the IR, this is somewhat like the Veneziano limit wherein $\frac{N_f}{N_c}$ is fixed (but, unlike \cite{metrics,MQGP}, $N_c,N_f\rightarrow\infty$ in the Veneziano limit  in, e.g., s\cite{Nitti_et_al}) as (in the IR) $\frac{N_f}{N_c}\sim\frac{N_f}{M}$ in \cite{metrics}. Note, the low energy or the IR is relative to the string scale. But these energies which are much less than the string scale, can still be much larger than $T_c$. Therefore, as regards the energy scales relevant to QCD, the number of colors can be tuned to three.

Thus, under a Seiberg-like duality cascade the $N\ D3$-branes are cascaded away and there is a finite $M$ left at the end corresponding to a strongly coupled IR-confining $SU(M)$ gauge theory; the finite temperature version of this $SU(M)$ gauge theory is what was considered in \cite{metrics}. So, at the end of the Seiberg-like duality cascade in the IR, the number of colors $N_c$ is identified with $M$, which in the `MQGP limit' can be tuned to equal 3. 

\item
{\bf Color-Flavor Enhancement of Length Scale in the IR}:  In the IR in the MQGP limit, with the inclusion of terms higher order in $g_s N_f$  in the RR and NS-NS three-form fluxes and the NLO terms in $N$ in the angular part of the metric, there occurs an IR color-flavor enhancement of the length scale as compared to a Planckian length scale in KS even for ${\cal O}(1)$ $M$, thereby showing that quantum corrections will be suppressed. This was discussed in \cite{NPB} and is summarized here. Using \cite{metrics}:
\begin{eqnarray}
\label{NeffMeffNfeff}
& & N_{\rm eff}(r) = N\left[ 1 + \frac{3 g_s M_{\rm eff}^2}{2\pi N}\left(\log r + \frac{3 g_s N_f^{\rm eff}}{2\pi}\left(\log r\right)^2\right)\right],\nonumber\\
& & M_{\rm eff}(r) = M + \frac{3g_s N_f M}{2\pi}\log r + \sum_{m\geq1}\sum_{n\geq1} N_f^m M^n f_{mn}(r),\nonumber\\
& & N^{\rm eff}_f(r) = N_f + \sum_{m\geq1}\sum_{n\geq0} N_f^m M^n g_{mn}(r),
\end{eqnarray}
wherein  the type IIB axion $C_0 =N_f^{\rm eff} \frac{\left(\psi - \phi_1-\phi_2\right)}{4\pi}$, the ten-dimensional warp factor h, disregarding the angular part, is given by:
\begin{eqnarray}
\label{h}
h & = & \frac{4\pi g_s}{r^4}\left[N_{\rm eff}(r) + \frac{9g_sM^2_{\rm eff}g_sN_f^{\rm eff}}{8\pi^2}\log r\right].
\end{eqnarray}
At the end of a Seiberg-like duality cascade, $N_{\rm eff}(r_0\in\rm IR)=0$ and writing $h \sim \frac{L^4}{r^4}$,  the length scale $L$ in the IR will be given by:
\begin{eqnarray}
\label{length-IR}
& & L\sim\sqrt{M}N_f^{\frac{3}{4}}\sqrt{\left(\sum_{m\geq0}\sum_{n\geq0}N_f^mM^nf_{mn}(\Lambda)\right)}\left(\sum_{l\geq0}\sum_{p\geq0}N_f^lM^p g_{lp}(\Lambda)\right)^{\frac{1}{4}}g_s^{\frac{1}{4}}\sqrt{\alpha^\prime}\nonumber\\
& & \equiv N_f^{\frac{3}{4}}\left.\sqrt{\left(\sum_{m\geq0}\sum_{n\geq0}N_f^mM^nf_{mn}(\Lambda)\right)}\left(\sum_{l\geq0}\sum_{p\geq0}N_f^lM^p g_{lp}(\Lambda)\right)^{\frac{1}{4}} L_{\rm KS}\right|_{\Lambda:\log \Lambda{<}{\frac{2\pi}{3g_sN_f}}},
\end{eqnarray}
which implies that  in the IR, relative to KS, there is a color-flavor enhancement of the length scale in the MQGP limit. Hence,  in the IR, even for $N_c^{\rm IR}=M=3$ and $N_f=2+1$ (for comparison with \cite{Petreczky-et-al}) upon inclusion of of $n,m>1$  terms in
$M_{\rm eff}$ and $N_f^{\rm eff}$ in (\ref{NeffMeffNfeff}), $L\gg L_{\rm KS}(\sim L_{\rm Planck})$ in the MQGP limit (\ref{MQGP_limit}) involving $g_s\stackrel{\sim}{<}1$, implying that {\it the stringy corrections are suppressed and one can trust supergravity calculations}. This is verified in {\bf 4.3} wherein it is explicitly shown (at low temperatures, i.e., $T<T_c$; we expect a similar result though even for high temperatures, i.e., $T>T_c$) that the ${\cal O}(R^4)$ corrections are suppressed as compared to the LO terms in the supergravity action.

\item
{\bf Gravity dual of the brane construct of \cite{metrics}}: The finite temperature on the gauge/brane side is effected in the gravitational dual via a black hole in the latter. Turning on of the temperature (in addition to requiring a finite separation between the $M\ D5$-branes and $M\ \overline{D5}$-branes to provide a natural scale above which one is in the UV) corresponds in the gravitational dual to having a non-trivial resolution parameter of the conifold. IR confinement on the brane/gauge theory side corresponds to having a non-trivial deformation of the conifold geometry in the gravitational dual.  The gravity dual is hence given by a  resolved warped deformed conifold wherein the $D3$-branes and the $D5$-branes are replaced by fluxes in the IR, and the back-reactions are included in the warp factor and fluxes.  
\end{itemize}

Hence, the type IIB model of \cite{metrics} make it an ideal holographic dual of thermal QCD because: (i) it is UV conformal (Landau poles are absent), (ii) it is IR confining, (iii) the quarks transform in the fundamental representation of flavor and color groups, and (iv) it  is defined for the full range of temperature - both low and high.

%The length scale on the gravity side in the IR will be given by:
%\begin{eqnarray*}
%label{length-IR}
%& & L\sim\sqrt{M}N_f^{\frac{3}{4}}\sqrt{\left(\sum_{m\geq0}\sum_{n\geq0}N_f^mM^nf_{mn}(\Lambda)\right)}\left(\sum_{l\geq0}\sum_{p\geq0}N_f^lM^p g_{lp}(\Lambda)\right)^{\frac{1}{4}}g_s^{\frac{1}{4}}\sqrt{\alpha^\prime}\nonumber\\
%& & \equiv N_f^{\frac{3}{4}}\sqrt{\left(\sum_{m\geq0}\sum_{n\geq0}N_f^mM^nf_{mn}(\Lambda)\right)}\left(\sum_{l\geq0}\sum_{p\geq0}N_f^lM^p g_{lp}(\Lambda)\right)^{\frac{1}{4}} L_{\rm KS},
%\end{eqnarray*}
%which implies that  in the IR, relative to KS, there is a color-flavor enhancement of the length scale. Hence,  in the IR, even for $N_c^{\rm IR}=M=3$ and $N_f=2$ (light flavors) upon inclusion of higher order terms in $M$ and $N_f$, $L\gg L_{\rm KS}(\sim L_{\rm Planck})$ in the MQGP limit involving $g_s\stackrel{\sim}{<}1$, implying that the stringy corrections are suppressed and one can trust supergravity calculations.

\item
{\bf The MQGP limit, Type IIA SYZ mirror \cite{syz} of \cite{metrics} and its M-theory uplift at intemediate gauge coupling}:
\begin{itemize}
\item
For constructing a holographic dual of thermal QCD-like theories, one would have to consider intemediate gauge coupling (as well as finite number of colors) $-$ dubbed as the `MQGP limit' in \cite{MQGP}. From the perspective of gauge-gravity duality, this necessitates looking at the strong-coupling/non-perturbative limit of string theory - M theory. The MQGP limit in \cite{MQGP} was defined as: 
\begin{equation}
\label{MQGP_limit}
g_s\stackrel{<}{\sim}1, M, N_f \equiv {\cal O}(1),\ N \gg1,\ \frac{g_s M^2}{N}\ll1.
\end{equation}
%\footnote{$M$ could also be ${\cal O}(1)$ such that the non-planar diagrams are still suppressed by $1/M$.} 

\item
The M-theory uplift of the type IIB holographic dual of \cite{metrics} was constructed in \cite{MQGP} by working out the SYZ type IIA mirror of \cite{metrics} implemented via a triple T duality along a local special Lagrangian (sLag) $T^3$ $-$ which could be identified with the $T^2$-invariant sLag of \cite{M.Ionel and M.Min-OO (2008)} with a large base ${\cal B}(r,\theta_1,\theta_2)$ (of a $T^3(\phi_1,\phi_2,\psi)$-fibration over ${\cal B}(r,\theta_1,\theta_2)$) \cite{NPB,EPJC-2}\footnote{Consider $D5$-branes wrapping the resolved $S^2$ of a resolved conifold geometry \cite{Zayas-Tseytlin}, which one knows, globally, breaks  SUSY (nicely explained in \cite{Franche-thesis}).  As in \cite{SYZ-free-delocalization}, to begin with, SYZ is implemented locally wherein the pair of $S^2$s are replaced by a pair of $T^2$s in the delocalized limit, and the correct T-duality coordinates are identified. Then,  when uplifting the mirror to M theory, it is found
 that a $G_2$-structure  can be chosen that is in fact, free, of the delocalization. For the SYZ mirror of the resolved warped deformed conifold uplifted to M-theory with $G_2$ structure worked out in \cite{MQGP}, the idea is precisely the same. Also note (as pointed out in Fig. 1), the type IIB/IIA $SU(3)$ structure torsion classes (in the MQGP limit and in the UV/UV-IR interpolating region), satisfy the same relationships as satisfied by corresponding supersymmetric conifold geometries \cite{Butti et al [2004]}.} Let us elucidate the basic idea. Let us consider the aforementioned $N$ D3-branes  oriented along $x^{0, 1, 2, 3}$ at the tip of conifold. Further, assume the $M\ D5$-branes to be parallel to these $D3$-branes as well as wrapping the vanishing $S^2(\theta_1,\phi_1)$. A single T-dual along $\psi$  yields $N\ D4$-branes wrapping the $\psi$ circle and $M\ D4$-branes straddling a pair of orthogonal $NS5$-branes. These pair of $NS5$-branes 
correspond to the vanishing $S^2(\theta_1,\phi_1)$ and the blown-up $S^2(\theta_2,\phi_2)$ with a non-zero resolution parameter $a$ - the radius of the blown-up $S^2(\theta_2,\phi_2)$. Two more T-dualities along $\phi_i$ and $\phi_2$, convert the aforementioned pairt of orthogonal $NS5$-branes into two orthogonal Taub-NUT spaces, the $N\ D4$-branes into $N$ color $D6$-branes and the $M$ straddling $D4$-branes also to $D6$-branes. Similarly, in the presence of the aforementioned $N_f$ flavor $D7$-branes (embedded holomorphically via the Ouyang embedding), oriented parallel to the $D3$-branes and ``wrapping" a non-compact four-cycle $\Sigma^{(4)}(r, \psi, \theta_1, \phi_1$), upon T-dualization yield $N_f$ $D6$-branes ``wrapping" a non-compact three-cycle  $\Sigma^{(3)}(r, \theta_1, \phi_2$). An uplift to M-theory of the SYZ type IIA mirror, will convert  the $D6$-branes to KK monopoles, which are variants of Taub-NUT spaces.  Therefore, all the branes are converted to geometry and fluxes, and one  ends up with M-theory on a $G_2$-structure manifold. Similarly, one may perform identical three T-dualities on the gravity dual on the type IIB side, which is a resolved warped-deformed conifold with fluxes,  to obtain another $G_2$ structure manifold, giving us the MQGP model of  \cite{MQGP,NPB}. 
\end{itemize} 
\end{enumerate}

%\section{Type IIB Dual of Large-$N$ Thermal QCD, its Delocalized type IIA SYZ Mirror and M-Theory Uplift in the `MQGP Limit'}
The type IIB brane construct, its type IIA mirror as well as the type IIB gravity dual, its SYZ IIA mirror gravity dual along with the M-theory uplift of the type IIA gravity dual are summarized in Fig. {\bf 1}. The $SU(3)/G_2$ structure torsion classes (which measure the deviation of a six/seven-fold from having $SU(3)/G_2$ holonomy) are denoted respectively by $W_{i=1,2,3,4,5}/W_{i=1,2,3,4}$ (with superscripts in the $G_2$-structure torsion classes denoting the respective dimensionalities) therein.

\begin{figure}
\begin{center}
\includegraphics[width=0.73\textwidth]{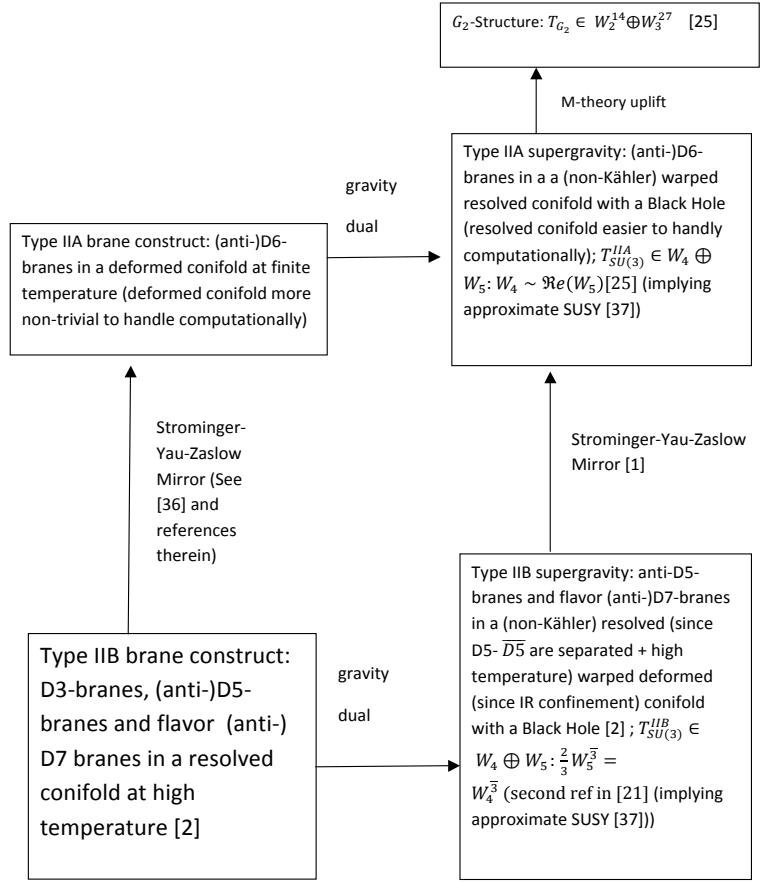}
\end{center}
\caption{Dasgupta et al's \cite{metrics} Type IIB brane construct and the gravity dual of large-N QCD at high temperature, and their (SYZ) mirrors}
\label{flowchart}
\end{figure}

%m

\section{Lattice-Compatible $T_c$ }

In this section after obtaining a lattice-compatible confinement-deconfinement phase transition temperature $T_c$ as a Hawking-Page phase transition at zero chemical potential, we obtain the temperature variation of the QCD trace anomaly from M theory, both, at large temperatures $T>T_c$ as well as low temperatures $T<T_c$.

The temperature at the horizon $r_h$ is given as under:
\begin{eqnarray}
\label{T}
& & T(r=r_h) =  \frac{\partial_rG^{\cal M}_{00}}{4\pi\sqrt{G^{\cal M}_{00}G^{\cal M}_{rr}}},\end{eqnarray}
which in the MQGP limit and utilizing the IR-valued warp factor $h(r,\theta_1,\theta_2)$:
\begin{eqnarray}
\label{h_small_r}
&& \hskip -0.45in h(r,\theta_1,\theta_2) = \frac{L^4}{r^4}\Bigg[1+\frac{3g_sM_{\rm eff}^2}{2\pi N}{\rm log}r\left\{1+\frac{3g_sN^{\rm eff}_f}{2\pi}\left({\rm
log}r+\frac{1}{2}\right)+\frac{g_sN^{\rm eff}_f}{4\pi}{\rm log}\left({\rm sin}\frac{\theta_1}{2}
{\rm sin}\frac{\theta_2}{2}\right)\right\}\Bigg],\nonumber\\
& & \hskip -0.45in L \equiv \left(4\pi g_s N\right)^{\frac{1}{4}},
\end{eqnarray}
will when evaluated near $\theta_1\sim N^{-\frac{1}{5}},\theta_2\sim N^{-\frac{3}{10}}$ [the specific values of small $\theta_{1,2}$ facilitated in \cite{NPB} construction of explicit $SU(3)/G_2$ structures respectively for the string/M theory duals; the EH and GHY terms receive their most dominant contributions near very small values of $\theta_{1,2}$; the same along with $\psi=0,2\pi,4\pi$ has the advantage of the decoupling of $M_5(x^{0,1,2,3},r)$ and $\tilde{M}_5(\theta_{1,2},\phi_{1,2},\psi)$], can be written out \cite{EPJC-2} in terms of $N, M$ and $N_f$. Now, we will take (as in \cite{EPJC-2}), the following form of the resolution parameter (the radius of the blown-up $S^2$ of the non-K\"{a}hler warped resolved conifold):
\begin{equation}
\label{a_rh}
a(r_h) = \left(\alpha + \gamma\frac{g_s M^2}{N} + \beta \frac{g_s M^2}{N}\log r_h\right)r_h.
\end{equation}

%For the purpose of comparison of $\frac{\eta}{s}$ with lattice/RHIC data for QGP and consequently be able to express
%$r_h$ in terms of $\tilde{t}\equiv \frac{T}{T_c}-1$,  in \cite{EPJC-2},
We will now see how to obtain a lattice-compatible $T_c$. We will implement the idea that in the absence of a chemical potential the confinement-to-deconfinement transition in the gravitational dual side, can be understood as a Hawking-Page first order phase transition from a thermal ($T<T_c$) gravity dual to the one consisting of a black hole ($T>T_c$) \cite{Witten-Tc}.   Inspired by \cite{metrics,EPJC-2}, the following type IIB dilaton ($\Phi$) profile will be assumed:
\begin{eqnarray}
\label{dilaton}
& (a) & T(r_h\neq0):\nonumber\\
& & e^{-\Phi} = \frac{1}{g_s} - \frac{N_f^{\rm eff}}{8\pi}\log(r^6 + a^2 r^4) - \frac{ N_f^{\rm eff}}{2\pi}\log\left(\sin\frac{\theta_1}{2}\sin\frac{\theta_2}{2}\right),\ r<{\cal R}_{D5/\overline{D5}},\nonumber\\
& & e^{-\Phi} = \frac{1}{g_s},\ r>{\cal R}_{D5/\overline{D5}};\nonumber\\
& (b) & T(r_h=0):\nonumber\\
& & e^{-\Phi} = \frac{1}{g_s} - \frac{3 N_f^{\rm eff}}{4\pi}\log r - \frac{N_f^{\rm eff}}{2\pi}\log\left(\sin\frac{\theta_1}{2}\sin\frac{\theta_2}{2}\right),\ r<\left|\mu_{\rm Ouyang}\right|^{\frac{2}{3}},\nonumber\\
& & e^{-\Phi} = \frac{1}{g_s},\ r>\left|\mu_{\rm Ouyang}\right|^{\frac{2}{3}},
\end{eqnarray}
wherein $N_f^{\rm eff}$ is the effective number of $D7$-branes (or the effective axionic charge) the Ouyang embedding parameter is defined via:
\begin{equation}
\label{Ouyang-definition}
\left(r^6 + 9 a^2 r^4\right)^{\frac{1}{4}} e^{\frac{i}{2}\left(\psi-\phi_1-\phi_2\right)}\sin\left(\frac{\theta_1}{2}\right)\sin\left(\frac{\theta_2}{2}\right) = \mu_{\rm Ouyang}.
\end{equation}
Hence, setting the Newtonian constant to unity,  performing a large-N expansion and then a large $R_{\rm UV}$-expansion,  for the thermal background ($r_h=0$) for which $r\in[r_0,R_{\rm UV}]$ where $r_0$ and $R_{\rm UV}$ are respectively the IR and UV cut-offs, the potential:
\begin{eqnarray}
\label{V_1}
& & \frac{V_1}{\Omega_5} = -\frac{1}{2}\int_{r=r_0}^{R_{\rm UV}} d^5x\sqrt{-g}e^{-2\Phi}\left(R - 2\Lambda\right) - \int_{r=R_{\rm UV}}d^4x\sqrt{-h}e^{-2\Phi}K,
\end{eqnarray}
where $\Omega_5\equiv\int d^5y\sqrt{g(\theta_{1,2},\phi_{1,2},\psi)}\delta\left(\theta_1-\frac{\alpha_{\theta_1}}{N^{\frac{1}{5}}}\right)\delta\left(\theta_2-\frac{\alpha_{\theta_2}}{N^{\frac{3}{10}}}\right)\delta\left(\psi - 2n\pi\right)(n=0/1/2, \alpha_{\theta_{1,2}}$ being ${\cal O}(1)$ numbers). Similarly for the black hole background, for which $r\in[r_h,R_{\rm UV}]$ the potential:
\begin{eqnarray}
\label{V_2}
& & \frac{V_2}{\Omega_5} = -\frac{1}{2}\int_{r=r_h}^{R_{\rm UV}} d^5x\sqrt{-g}e^{-2\Phi}\left(R - 2\Lambda\right) - \int_{r=R_{\rm UV}}\sqrt{-h}d^4xe^{-2\Phi}K
\end{eqnarray}
was worked out in \cite{EPJC-2}. Counter terms involving $\int_{R_{\rm UV}}\sqrt{-h^{\rm Thermal/BH}}$ need to be subtracted from $V_{1,2}$ to render them UV-finite and it was shown in \cite{EPJC-2} that assuming
${\cal R}_{D5/\overline{D5}} = \sqrt{3}a$, $|\mu_{\rm Ouyang}|^{\frac{2}{3}}=\delta(\equiv {\cal O}(1))\times r_0$ and assuming an IR-valued $r_h, r_0$, $(V_2 - V_1)^{\rm UV-finite} =0$ yields:
\begin{equation}
\label{r0rh}
r_0  =  r_h\sqrt[4]{\left|\frac{9 \alpha ^4-1}{2(\delta^{\frac{8}{3}}-1)}\right|} + {\cal O}\left(\frac{1}{\log N}\right).
\end{equation}
Now, as we was shown in \cite{Sil+Yadav+Misra-glueball}, the lightest $0^{++}$ scalar glueball mass is given by:
\begin{equation}
\label{glueball-mass-i}
m_{0^{++}} \approx \frac{4 r_0}{L^2}.
\end{equation}
Now, lattice calculations for $0^{++}$ scalar glueball mass \cite{SU3_lattice_glueball_masses}, yield the lightest mass to be around $1,700$ MeV. Hence, to make contact with lattice results, using (\ref{glueball-mass-i}),  $\frac{r_0}{L^2}$ is replaced by $\frac{1,700}{4}$ (in units of MeV) to yield:
\begin{equation}
\label{Tc}
T_c = \left.\frac{m_{\rm glueball}\left(1 + \frac{3\alpha^2}{2}\right)}{2^{\frac{7}{4}}\pi\sqrt[4]{\left|\frac{9 \alpha ^4-1}{2(\delta^{\frac{8}{3}}-1)}\right|}}\right|_{\alpha=0.6,\delta=1.008}
=150 \ {\rm MeV},
\end{equation}
which is what is expected from lattice QCD in the heavy-quark-mass limit. Let us elaborate more. One should note that in our gravity dual as proposed in \cite{Witten-Tc}, the Hawking-Page phase transition occurs at zero baryon chemical potential $\mu_C$ and one expects a smooth cross-over for a non-zero $\mu_C$ above a critical value. However, it is exactly the opposite of what one generically expects from (lattice) QCD. But, as explained in \cite{He:2013qq}, our holographic gravity dual computation can still be justified in the heavy quark limit wherein the first order phase transition at $\mu_C=0$ becomes a cross-over for $\mu_C\neq0$. Let us explain how the heavy quark-mass limit is implied in our calculations and hence ensure compatibility with the lattice results of \cite{He:2013qq}. We assume that in the type IIB dual (whose uplift via the type IIA SYZ mirror is the M theory dual we are working with in Section {\bf 2}), all $D7$-branes have been identically embedded; in other words, in the type IIB brane picture, the quarks corresponding to the $D3-D7$ strings, are either all light or are all heavy - this will be determined by the modulus of the Ouyang embedding parameter. The reason is that the (modulus of the) Ouyang embedding parameter $\mu_{\rm Ouyang}$ has the physical interpretation that $|\mu_{\rm Ouyang}|^{\frac{2}{3}}$ gives essentially the mass of the fundamental quarks arising from the $D3-D7$ strings in the type IIB string theory dual as constructed in \cite{metrics}. Now, in \cite{NPB} and the first reference in \cite{EPJC-2},  it was shown that : $|\mu_{\rm Ouyang}| \sim {r_h^{-\alpha}}, \alpha>0$. Further, the horizon radius $r_h$ was estimated in the third reference in \cite{EPJC-2} to be:
\begin{equation}
\label{r_h}
r_h \sim \exp \left[-\frac{1}{3(6\pi)^{\frac{1}{3}} \left(g_s N_f\right)^{\frac{2}{3}}\left(\frac{g_sM^2}{N}\right)^{\frac{1}{3}}}\right],
\end{equation}
 implying a very small $r_h$ and hence a large $|\mu_{\rm Ouyang}|$ in the large-$N$ `MQGP limit' of \cite{MQGP}. So, the quark mass indirectly enters
our M theory computations via $r_h$, which in the MQGP limit automatically implies considering the heavy-quark-mass limit.

The lattice calculations of \cite{He:2013qq} we have compared with in this paper, have $N_f=2(u,d)+1(s)$ wherein $m_{u/d}=\frac{m_s}{20}$. So, we can safely consider $u/d$ quarks to be light and $s$ quark to be heavy. The trace anomaly and other thermodynamic quantities are  seen in \cite{He:2013qq} to be insensitive to $m_{u/d}$ for $T>300$ MeV; hence, at least for high temperatures it is acceptable if one assumes that only the heavy/strange quark contributes to  the trace anomaly.

%What this requires is to consider the GHY (boundary) supergravity action as well as the DBI action for the flavor branes in the string theory dual.  This will be the topic of follow-up work. However, the following observations can readily be made right now itself. 

 From the evaluation of the baryon chemical potential $\mu_C$ in appendix {\bf A} we see that the $|\mu_{\rm Ouyang}|\gg1$-limit corresponding to the heavy-quark limit of $\mu_C$ or $|\mu_{\rm Ouyang}|\ll1$-limit corresponding to the light-quark limit of $\mu_C$  which would imply that all $N_f$ flavors are respectively equally heavy or light, yields: $\mu_C\rightarrow0$. If one assumes that all quarks are s-like, in other words, ``heavy'', we can also obtain at least a qualitative agreement  between (\ref{muC-2}) and (\ref{muC-3}) and the $\left(\frac{\mu_C}{T}\right)^2$-vs-$\frac{m_s}{T}$ cross-section of Fig. 16 of  \cite{Fromm:2011qi}.

Similarly, from the evaluation of the DBI action on the flavor $D7$-branes in appendix ${\bf A}$,  one notices that in the light-quark-mass limit, effected by $|\mu_{\rm Ouyang}|\ll 1$-limit, the UV-finite part of the DBI action (i.e. the part that remains finite in the large-UV-cutoff limit) vanishes. In the heavy-quark-mass limit effected by $|\mu_{\rm Ouyang}|\gg 1$-limit, using (\ref{r_h}), one sees that there  is no large-$N$-finite contribution that survives from the UV-finite part of the DBI action. Therefore, in the light- or heavy-quark mass limit wherein $\mu_C=0$, the UV-large-N finite contribution effectively arises only from the supergravity action alone and not the DBI action; as shown above, the former yields a first order Hawking-Page phase transition.  Hence, like the famous ``Columbia plot" of phase transition/cross-over in $N_f=2+1$ QCD, we have a phase transition in the light/heavy quark-mass limit corresponding to vanishing baryon chemical potential. 

\section{QCD Trace Anomaly from M Theory}

In this section, we will compute the QCD trace anomaly hologarphically from M theory. This computation is divided into two subsections - {\bf 4.1} addresses the large temperature regime, i.e.,
$T>T_c$, and {\bf 4.2} addresses the low temperature regime, i.e., $T<T_c$.

The UV-finite part of the $D=11$ supergravity action is given as under:
\begin{eqnarray}
\label{D11SUGRAS}
& & \hskip -0.5in {\cal S}_{D=11} = \frac{1}{2\kappa_{11}^2}\Biggl[\int_{M_{11}}\sqrt{G^{\cal M}}R + \int_{\partial M_{11}}\sqrt{h}K -\frac{1}{2}\int_{M_{11}}\sqrt{G^{\cal M}}G_4^2
-\frac{1}{6}\int_{M_{11}}C_3\wedge G_4\wedge G_4\nonumber\\
& & \hskip -0.5in  + \frac{\left(4\pi\kappa_{11}^2\right)^{\frac{2}{3}}}{{(2\pi)}^4 3^2.2^{13}}\Biggl(\int_{\cal{M}} d^{11}\!x \sqrt{G^{\cal M}}\left(J_0-\frac{1}{2}E_8\right) + \int C_3 \wedge X_8 + \int t_8 t_8 G^2 R^3 + \cdot\Biggr)\Biggr] - {\cal S}^{\rm ct},
\end{eqnarray}
where $G^{\cal M}$ is the determinant of the $D=11$ metric, $h$ is the same restricted to $r$ fixed at the UV cut-off, $R$ is the $D=11$ Ricci scalar, $K$ is the extrinsic curvature with $\sqrt{h}K$ being the Gibbons-Hawking-York(GHY) surface term, $G_4=dC_3$, $C_3$ being the $D=11$ three-form potential, $\kappa_{11}$ being the $D=11$ Newtonian constant, and  $J_0, E_8, X_8, t_8$ and $G^2R^4$ are defined in Appendix {\bf B}; the ellipsis in (\ref{D11SUGRAS}) denoting terms in \cite{G^2R^3} other than the one explicitly mentioned in (\ref{J+E_8+X_8}) -   and the counter-term ${\cal S}_{\rm ct}$ is added such that the Euclidean action ${\cal S}_{D=11}$ is finite.   

To evaluate the boundary trace anomaly consider the following infinitesimal  Weyl transformation \cite{Thesien_et_al_holographic_trace_anomaly, Freedman+Proeyen-SUGRA}:
\begin{equation}
\label{infinitesimal-Weyl-transf}
\delta r = -2 r \delta\sigma(x),\ \ \delta h_{mn} (x) = 2 \delta\sigma(x)h_{mn}(x),
\end{equation}
where $\delta \sigma(x)$ is a local infinitesimal Weyl transformation parameter, and  $(m,n)\neq r, r\sim r_{\rm UV}\gg r_h$, the UV cut-off. The trace anomaly is then given by:
\begin{equation}
\label{trace-T}
{\cal T}^m_{\ m} = -\frac{1}{\sqrt{h}}\frac{\delta \left( S_{\rm on-shell} + S_{\rm ct} \right)}{\delta\sigma(x)}.
\end{equation}

Let us now discuss the computation of the conformal trace anomaly via the application of (\ref{infinitesimal-Weyl-transf}) - (\ref{trace-T}), separately for $T>T_c$ ({\bf 4.1}) and $T<T_c$ ({\bf 4.2}).

\subsection{High Temperatures ($T>T_c$)}

In this subsection, we  will evaluate the trace anomaly for large temperatures, i.e., $T>T_c$, and compare our results with the lattice results in \cite{Petreczky-et-al}\footnote{For previous bottom-up holographic computation-compatible lattice results, see, e.g., \cite{Panero-PRL, Lucini+Panero-Phys-Rept}; see \cite{Bringoltz+Teper} for previous large-$N$ lattice results for the trace anomaly.}. The upshot of this subsection is that it is only the counter-term used to cancel the UV divergence generated from the GHY boundary term that contributes to the trace of the energy momentum tensor. The variation of the aforemetioned counter term, with respect to the scalar appearing in the Weyl scaling of the radial coordinate and the metric along the other non-radial directions, generates the same temperature-dependent contribution as from the extrinsic curvature itself.

%One notes that under $\delta h_{\mu\nu} = - 2\delta \sigma(x)g_{\mu\nu}, (\mu,\nu)\in x^{0,1,2,3}$ alone, $\delta K=0$. Alternatively, as the same is equivalent to \cite{Thesien_et_al_holographic_trace_anomaly}: $x^\mu = x^\prime\ ^\mu + a^\mu(x^\prime,u^\prime)$, where $a^\mu(x,u) = \frac{L^2}{2}\int_0^u du^\prime g^{\mu\nu}(x,u^\prime)\partial_\nu\delta\sigma(x), u=\frac{r_h}{r}$. As there is no dependence of the (renormalized) supergravity Lagrangian on $x^\mu$, the former is invariant under the variation of the latter. So, one need worry only about $\delta K$ under $\delta r = - 2 \delta \sigma (x) r$. 
Now, on-shell: 
\begin{eqnarray}
\label{Einstein_EOM_metric-D=11}
& & R_{MN} - \frac{g_{MN}}{2}R - \frac{1}{12}\left(G_{MPQR}G_N^{\ \ PQR} - \frac{g_{MN}}{8}G^2\right) = -\beta\frac{1}{\sqrt{-g}}\frac{\delta}{\delta g^{MN}}\left[\sqrt{-g}\left(J_0 - \frac{E_8}{2}\right)\right],\nonumber\\
& & 
\end{eqnarray}
$\beta$ defined just above (\ref{EOM-C_3}). As will be shown in {\bf 3.3}, the RHS is sub-dominant as compared to the LHS of (\ref{Einstein_EOM_metric-D=11}). Hence, $R_{D=11}\sim G^2>0$, and one can write: $\int \sqrt{G^{\cal M}}R \sim \Lambda\int \sqrt{G^{\cal M}}$ where the flux-generated cosmological constant $\Lambda$ is given by: $\Lambda\int \sqrt{G^{\cal M}} \sim \int G_4\wedge *G_4$. 
One can show \cite{Sil+Misra+Yadav_Glueball}:
\begin{eqnarray}
\label{Lambda}
& & \hskip -0.3in\int \left(G_4\wedge *G_4 \approx \right)\sqrt{G^{\cal M}}\left|H_3^{\rm IIA}\wedge A^{\rm IIA}\right|^2
\sim \int \left.\sqrt{G^{\cal M}}\left|H_3^{\rm IIA}\wedge A^{\rm IIA}\right|^2\right|_{\rm Ouyang\ embedding_{\rm UV}}\nonumber\\
& & \hskip -0.3in \sim \int \sqrt{G^{\cal M}}\left|H_3^{\rm IIA}\wedge A^{\rm IIA}\right|^2\delta\left(\theta_1-\frac{\alpha_{\theta_1}}{N^{\frac{1}{5}}}\right)\delta\left(\theta_2-\frac{\alpha_{\theta_2}}{N^{\frac{3}{10}}}\right)\delta\left(\psi - 2n\pi\right)
%\sim \int dr \frac{a^4 {g_s}^{9/4} M^4 {N_f}^3 r^3 \log (r)}{\sqrt[4]{N} \log ^2(N) \alpha _{\theta _1}^{11} \alpha _{\theta _2}} 
\nonumber\\
& & \hskip -0.3in\sim \frac{a^4{g_s^{\rm UV}\ }^{13/4} M_{\rm UV}^4 N^{3/4} {N_f^{\rm UV}}\ ^{5/3} \log \left(\frac{R_{\rm UV}^4}{e}\right)}{\alpha _{\theta _1}^{11} \alpha _{\theta _2}
   \log ^{\frac{10}{3}}(N) \sqrt{1-\frac{{r_h}^4}{R_{\rm UV}^4}}}\int_{r=R_{\rm UV}}d^4x\sqrt{g_{M_4(x^{0,1,2,3})}},\nonumber\\
& & 
\end{eqnarray}
where $ N_f^{\rm UV}/M_{\rm UV} $ are average values of $M_{\rm eff}(r), N_f^{\rm eff}(r)$ in the UV, $H_3^{\rm IIA}$ being the SYZ type IIA mirror/tripe T-dual NS-NS three-form and $A$ is the type IIA RR one-form generated from the SYZ/triple T-dual of $F_{1,3,5}^{\rm IIB}$ of \cite{metrics}. In the UV (to ensure conformality), $ M_{\rm UV}, N_f^{\rm UV}\approx0$. The contribution to the action from $\int G_4\wedge*G_4$ is further suppressed by the (small) resolution parameter-dependent factor $a^4$.

As (the extrinsic curvature) $K=\frac{\sqrt{G^{{\cal M}\ rr}}}{2}\frac{\partial_r {\rm det}\ h_{mn}}{{\rm det}\ h_{mn}}, (m,n)\neq r$, where $h_{mn}=G^{\cal M}_{mn}(r=R_{\rm UV}=$ UV cut-off), one can show that 
%\footnote{One needs to impose that $r\rightarrow R_{\rm UV}: \log R_{\rm UV} = \frac{\log N}{3}$.}:
\begin{eqnarray}
\label{K_same_D=5_D=11}
K_{\partial M_5(x^{0,1,2,3}; r=R_{\rm UV})}\sim K_{\partial M_{11}(x^{0,1,2,3},\theta_{1,2},\phi_{1,2},\psi,x^{10}; r=R_{\rm UV})} \sim \left(\frac{1}{L  \sqrt[3]{{N_f}^{\rm UV}}\log N}\right)\sqrt{1 - \frac{r_h^4}{R_{\rm UV}^4}}.\nonumber\\
& & 
%\sim -\frac{r_h^4}{L R_{\rm UV}^4},
\end{eqnarray}
%where $L = \left(4 \pi g_s N\right)^{\frac{1}{4}}$ in $\alpha^\prime=1$-units.  
Therefore, effectively  we have dimensionally reduced the M-theory conformal anomaly to a $D=5$ holographic conformal anomaly. Near $\left(\theta_1,\theta_2\right)\sim \left(N^{-\frac{1}{5}},N^{-\frac{3}{10}}\right)$, one can show:
\begin{eqnarray}
\label{On_Shell_S}
& & S_{\rm on-shell} + S_{\rm ct} \nonumber\\
& &   = S_{\rm flux\ \Lambda} + S_{\rm GHY} \nonumber\\
& &  -\alpha_{\rm ct} \frac{{g_s^{\rm UV}}\ ^{7/4} M_{\rm UV} N^{19/20} {N_f^{\rm UV}}\ ^{5/3} \log ^{\frac{5}{3}}(N) \sqrt{1-\frac{{r_h}^4}{R_{\rm UV}^4}}
   \log (R_{\rm UV})}{\alpha _{\theta _1}^3 \alpha _{\theta _2}^2}\int_{r=R_{\rm UV}}d^4x\sqrt{g_{M_4(x^{0,1,2,3})}(r=R_{\rm UV})},\nonumber\\
& & 
\end{eqnarray}
$\alpha_{\rm ct}$ being an appropriate constant. As $S_{\rm flux\ \Lambda} + S_{\rm GHY}$ is invariant under (\ref{infinitesimal-Weyl-transf}), therefore, from (\ref{trace-T}) it is only the $R_{\rm UV}$-dependent factors in the counter term required to cancel the UV-divergent contribution arising from the GHY boundary term (very similar to the example in [section 23.11.2 of] \cite{Freedman+Proeyen-SUGRA}) that contributes to the trace of the energy momentum tensor and yields:
\begin{eqnarray}
\label{Trace_T>T_c}
 \left({\cal T}^\mu_{\ \mu}\right)_{\rm grav}& \sim & -{\alpha _{\rm ct}}\frac{\sqrt{1-\frac{{r_h}^4}{R_{\rm UV}^4}}+\frac{2 {r_h}^4 \log (R_{\rm UV})}{R_{\rm UV}^4
   \sqrt{1-\frac{{r_h}^4}{R_{\rm UV}^4}}}}{L \sqrt[3]{{N_f^{\rm UV}}} \sqrt[3]{\log (N)} \log (R_{\rm UV})}\nonumber\\
& & \sim-{\alpha _{\rm ct}}\frac{1}{L \sqrt[3]{{N_f}^{\rm UV}} \sqrt[3]{\log (N)}}\Biggl[\frac{1}{\log R_{\rm UV}} + 2 \left(\frac{ r_h}{R_{\rm UV}}\right)^4
 + {\cal O}\left( \left(\frac{ r_h}{R_{\rm UV}}\right)^8\right)\Biggr].\nonumber\\
& & 
\end{eqnarray}
%where the last equivalence as stated above is up to a temperature-independent boundary cosmological constant which can be absorbed into a 5D boundary counter term. 
Now, if $ M_{\rm UV},  N_f^{\rm UV}, \alpha_{\theta_{1,2}}$ are chosen such that the contribution from $ S_{\rm flux\ \Lambda} $ exactly cancels off \\ $\frac{\alpha _{\rm ct}}{L \sqrt[3]{{N_f}^{\rm UV}} \sqrt[3]{\log (N)} \log (R_{\rm UV})}$,
%\footnote{The contribution of (\ref{Lambda}) to the conformal/trace anomaly, using (\ref{trace-T}), is $\frac{a^4 {g_s^{\rm UV}}\ ^{5/4} M^3 \alpha _{\theta _2}}{N^{9/20} \sqrt[3]{{N_f^{\rm UV}}} \alpha _{\theta _1}^8 \log ^{\frac{16}{3}}(N)\sqrt{1-\frac{{r_h}^4}{R_{\rm UV}^4}} \log (R_{\rm UV})}$. }
i.e.:
\begin{equation}
\label{boundary-cc}
a=\frac{\sqrt[20]{N} \alpha _{\theta _1}^2 \sqrt[4]{\alpha _{{\rm ct}}} \log ^{\frac{5}{4}}(N)}{{{\cal O}(1)g_s^{\rm UV}}\ ^{3/8} M_{\rm UV}^{3/4}
   \sqrt[12]{{N_f^{\rm UV}}} \sqrt[4]{\alpha _{\theta _2}}}.
\end{equation}
The temperature dependence on the right hand side of (\ref{boundary-cc}), using (\ref{Ouyang-definition}), appears via $\alpha_{\theta_{1,2}}$ as  $R_{\rm UV}^{\frac{3}{2}}\frac{\alpha_{\theta_1} \alpha_{\theta_2}}{\sqrt{N}} \sim |\mu_{\rm Ouyang}|$ [given the temperature dependence of the Ouyang embedding parameter$\mu_{\rm Ouyang}$ (See \cite{NPB} and the first reference in \cite{EPJC-2})].  Alternatively, one could instead consider $\left(\tilde{\cal T}^\mu_{\ \mu}\right)_{\rm grav}\equiv \left({\cal T}^\mu_{\ \mu}\right)_{\rm grav} + \frac{\alpha _{\rm ct}}{L \sqrt[3]{{N_f}^{\rm UV}} \sqrt[3]{\log (N)} \log (R_{\rm UV})} $ for comparison with \cite{Petreczky-et-al}. We thus see that $\left({\cal T}^\mu_{\ \mu}\right)_{\rm grav}$ is given by the temperature/$r_h$-dependent contribution of $\sim K_{\partial M_5(x^{0,1,2,3}; r=R_{\rm UV})}$.
One hence concludes that  using dimensional consideration and noting that string/M theory uses a mostly positive Minkowskian signature whereas field theory uses a mostly negative Minkowskian signature, $\left({\cal T}^\mu_{\ \mu}\right)_{\rm grav}\rightarrow -\left({\cal T}^\mu_{\ \mu}\right)_{\rm FT}\equiv -{\cal T}^\mu_{\ \mu}$ (this notation will also be used in {\bf 3.2}):
\begin{equation}
\label{anomaly-expression}
\frac{{\cal T}^{\mu}_{\ \mu}}{T_c^4} (T>T_c) \sim\frac{1}{L \sqrt[3]{{N_f}^{\rm UV}} \sqrt[3]{\log (N)}}\left(\frac{ r_h}{R_{\rm UV}}\right)^4.
\end{equation}

\begin{figure}[h]
 \begin{center}
%\begin{center}
 \includegraphics[scale=0.7]
 %[height= 21cm,width=+15cm]
 {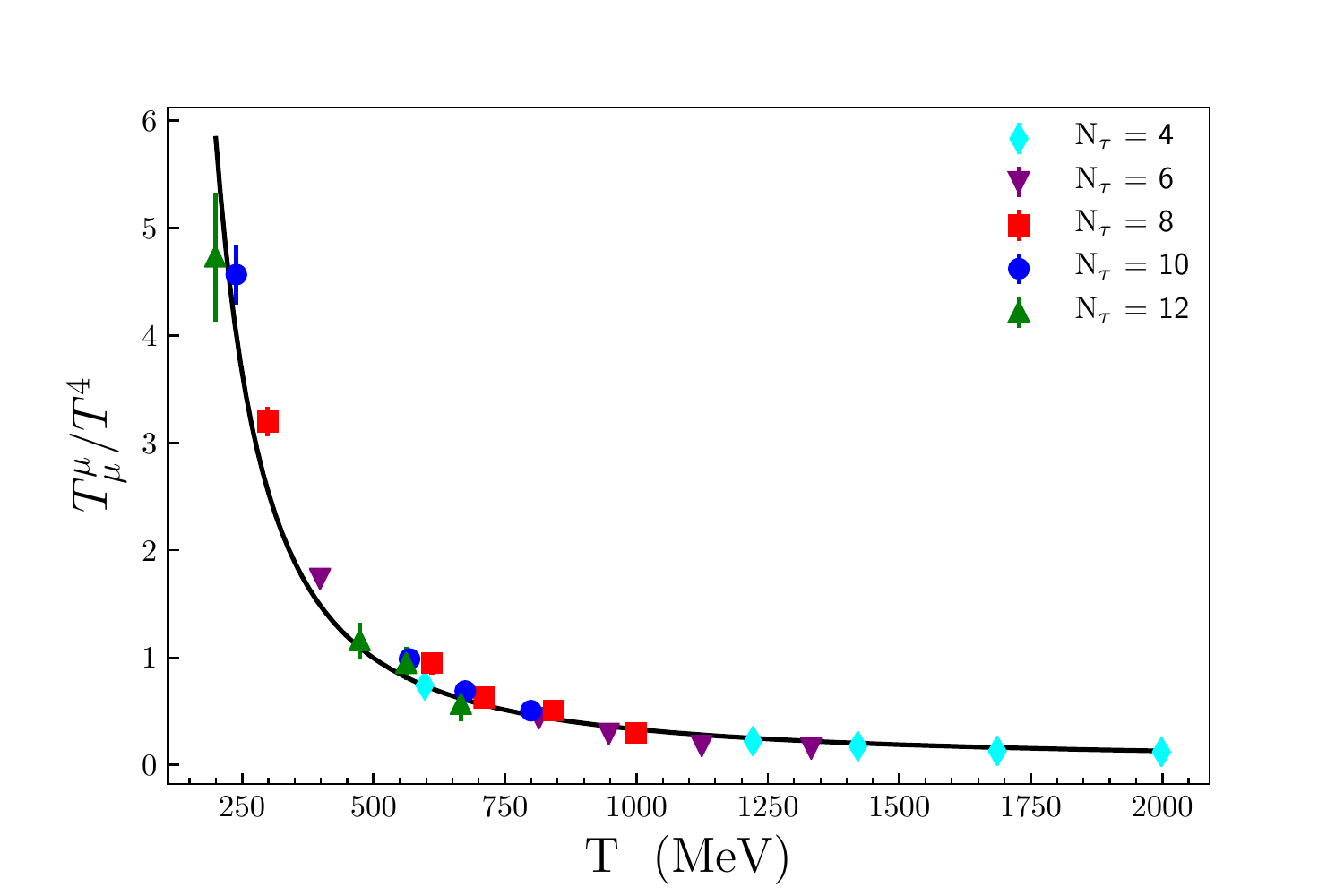}
 \end{center}
\caption{The scaled trace anomaly, $\frac{T^{\mu}_{\ \mu} }{T^4}$, plotted as a function of temperature for the high temperature region. The points represent lattice QCD results (Fig. 2 of Reference \cite{Petreczky-et-al}). The different $N_\tau$ values refer to the number of lattice cells in the imaginary time direction. The full curve is the fit obtained with the approach described in {\bf 2.1}. } 
\label{Lattice_highT}
\end{figure}
Matching $L \left(\frac{R_{\rm UV}}{R}\right)^4\sqrt[3]{{N_f}^{\rm UV}} \sqrt[3]{\log (N)}\frac{{\cal T}^{\mu}_{\ \mu}}{T^4}\equiv \frac{T^{\mu}_{\ \mu}}{T^4} = \frac{r_h^4}{R^4\tilde{t}^4}$ ($R$ being a scaling factor to match the lattice results of \cite{Petreczky-et-al}, and $\tilde{t}\equiv\frac{T}{T_c}$) with the data points of \cite{Petreczky-et-al} yields: 
\begin{equation}
\label{M-theory-large-T-anomaly/T^4}
\frac{T^{\mu}_{\ \mu}}{T^4} = \frac{\gamma }{\left({\cal P}{\cal L}[\omega  T]\right)^4},
\end{equation}
where $\gamma$ and $\omega$ are numerical constants, and ${\cal P L} (z)$ is the so-called Product log function, also known as the  Lambert W function \cite{Corless:1996zz} \footnote{As an example, requiring from \cite{Petreczky-et-al}, $\frac{T^{\mu}_{\ \mu}}{T^4}=4.73291$ at $T=199.184$ MeV, and
 $\frac{ T^{\mu}_{\ \mu}}{T^4}=0.120615$ at
$T=1998.47$ MeV implies that for $g_s=0.3, N=100, M=N_f=3$:
\begin{eqnarray*}
\label{R-lattice}
& & R = \frac{70501.1}{\Biggl[\left(\sum_{n=0}^2\alpha_n\beta^n\right) \left(\sum_{m=0}^2\gamma_m\beta^m\right)\Biggr]^{\frac{1}{4}} {\cal P}{\cal L}\left(\mathbb{D}\right)};\nonumber\\
   & & \beta = 43.5944,
\end{eqnarray*}
where $\alpha_2=\gamma_2=1$ and $\alpha_{n=0,1}$ and $\gamma_{m=0,1}$ are known numerical constants, and:
\begin{eqnarray*}
\label{mathbbD}
& & \mathbb{D} \equiv \frac{12449
   \left(\frac{520589}{39898748}\right)^{\mathbb{N}_1}
   3^{\mathbb{N}_2} 5^{\mathbb{N}_3}}{\cal D};
\end{eqnarray*}
\begin{eqnarray*}
\label{defs}
& & \mathbb{N}_1 \equiv \frac{\frac{2602633}{10073826406}-\frac{471991 \beta
   }{8537539578}}{\cal D},\nonumber\\
& & \mathbb{N}_2 \equiv -\mathbb{N}_1-1,\nonumber\\
& & \mathbb{N}_3 \equiv 4\mathbb{N}_1 - 5,\nonumber\\
& & {\cal D} \equiv \frac{891009 \beta
   }{136759244780}-\frac{87826}{69424325905}.
\end{eqnarray*}
This yields the following M-theory result for the anomaly: 
\begin{equation*}
\label{M-theory-large-T-anomaly/T^4}
\frac{T^{\mu}_{\ \mu}}{T^4} = \frac{3.449}{\left({\cal P}{\cal L}[0.0117 T]\right)^4}.
\end{equation*}}.   
A very good global chi-squared minimization fit which includes the high temperature lattice results is shown on Figure \ref{Lattice_highT}, and produces
\begin{eqnarray}
\gamma =  3.08 \pm 0.33, \omega =0.0100 \pm 0.0006\ ,
\end{eqnarray}
where the uncertainties are obtained from the diagonal elements of the covariance matrix. 

\subsection{Examining the low temperature region $T<T_c$}

For low temperatures, i.e. $T<T_c$, it is the thermal background ($r_h=0$) that is energetically preferred over the black-hole gravitational dual. Therein, the GHY action too becomes UV divergent in this limit.  Further, from (\ref{anomaly-expression}),  we see that one does not obtain any temperature-dependent contribution to the trace anomaly from the extrinsic  curvature $K$ upon setting $r_h=0$. Now, unlike for temperatures $T>T_c$ - corresponding to a gravitational dual with a black hole -  wherein the temperature is constrained to be given in terms of $r_h$,  for low temperatures ($T<T_c$), the temperature is a free parameter. For the thermal case, we will continue to use a resolved warped conifold in the type IIA mirror of \cite{MQGP}, assuming the existence of an $r_h$-independent bare resolution parameter $a_0$ in the resolution parameter $a$ guaranteeing the separation of the $M$ $D5-\overline{D5}$ branes ${\cal R}_{D5/\overline{D5}} \equiv \sqrt{3}a$ (as for resolved conifolds radial distances exceeding $\sqrt{3}a$, are taken to be large). This on the supergravity side, provides a natural scale which will be the boundary common to the IR-UV interpolating region and the UV. The horizon radius $r_h$ is the IR cut-off for high temperatures ($T>T_c$); for low temperatures ($T<T_c$) the IR cut-off is denoted by $r_0$ (See above (\ref{V_1})).  
%m(Analogous to $T>T_c$) For $T<T_c$ assuming $r_0=r_0(a_0) = a_0 + ...$, we will declare henceforth that  $T=T(a_0)\sim a_0$.  
From (\ref{r0rh}), one sees that $r_0$ and $r_h$ are proportional to each other; from (\ref{r_h}) one notes that $r_h$, in the MQGP limit, is very small. Hence, for $a_0\neq0$, it is possible to arrange: ${\cal R}_{D5/\overline{D5}}\gg r_0$.

Consider the flux term: $\int G_4\wedge * G_4$ in the $r_h=0$ limit. Using the results of \cite{MQGP, Sil+Misra+Yadav_Glueball}, one sees that:
\begin{eqnarray}
\label{flux1}
& & \int G_4\wedge *G_4 = \int_{\rm IR} G_4\wedge *G_4 + \int_{\rm UV}G_4\wedge *G_4,
\end{eqnarray}
where:
\begin{eqnarray}
\label{flux2}
& & {\rm IR}:\  \int G_4\wedge *G_4 \sim \frac{g_s^{\frac{9}{4}} M^4_{\rm IR} N_{f\ {\rm IR}}^3 }{N^{\frac{1}{4}}} \int_{r_0}^{\sqrt{3}a}dr\ a^6 \frac{d}{dr}\left(r^2 \log r\right),\nonumber\\
& & {\rm UV}:\ \int G_4\wedge *G_4 \sim \frac{g_s^{\frac{9}{4}} M^4_{\rm UV} N_{f\ {\rm UV}}^3}{N^{\frac{1}{4}}} \int_{\sqrt{3}a}^{R_{\rm UV}}dr\ a^4 \frac{d}{dr}\left(r^4 \log r\right).
\end{eqnarray}
The fact, that on-shell, the only contribution from the flux terms in the MQGP limit arises from $\int_{\partial M_{11}}C_3\wedge*G_4$, which is what (\ref{flux2}) in fact is, can be justified as follows. Defining $\beta l_p^6\equiv  \frac{\left(4\pi\kappa_{11}^2\right)^{\frac{2}{3}}}{{(2\pi)}^4 3^2.2^{13}}$, the $C_3$ EOM is:
\begin{equation}
\label{EOM-C_3}
d*G_4 = \frac{1}{2} G_4\wedge G_4 + \beta l_p^6 X_8.
\end{equation}
 The flux-dependent terms in (\ref{D11SUGRAS}), disregarding the $t_8t_8G^2R^3$ term - see the last paragraph of Subsection {\bf 2.2} -  can be rewritten as: 
\begin{eqnarray}
\label{flux_D=11-SUGRA}
\frac{1}{2}\left(\int_{\partial M_{11}}C_3\wedge*G_4 + \int_{M_{11}}C_3\wedge d*G_4\right)
+ \frac{1}{6}\int_{M_{11}}G_4\wedge G_4\wedge C_3 - \beta l_p^6\int_{M_{11}}C_3\wedge X_8.
\nonumber\\
& & 
\end{eqnarray}
On-shell, using (\ref{EOM-C_3}), one obtains:
\begin{eqnarray}
\label{Flux-On-Shell}
\frac{1}{2}\int_{\partial M_{11}}C_3\wedge*G_4 + \frac{5}{12}\int_{M_{11}}G_4\wedge G_4\wedge C_3 - \frac{\beta l_p^6}{2} \int_{M_{11}}C_3\wedge X_8.
\end{eqnarray}
Now, it was shown in \cite{MQGP} that $\int G_4\wedge G_4\wedge C_3 =\int C_3\wedge  X_8 = 0$ implying the assertion.

For $r_h=0$, one can arrange $(\int R *1\sim)\int_{\rm UV} G_4\wedge *G_4$ to cancel off $\int \sqrt{h}K(r=R_{\rm UV})$. This will be effected via: 
\begin{equation}
\label{GHY-cancel-FluxM_UV}
a = 
\frac{1}{{\cal O}(1)}\frac{N^{\frac{1}{20}}\left(\log N\right)^{\frac{5}{4}} \alpha_{\theta_1}^2}{\left(g_s^{\rm UV}\right)^{\frac{3}{8}}M_{\rm UV}^{\frac{3}{4}}\alpha_{\theta_2}^{\frac{1}{4}}},
\end{equation}
 with the understanding that $N\sim{\cal O}(10^2)$.

Thus, effectively, it is not the boundary in the UV at $r=R_{\rm UV}$ but the boundary common to the IR-UV interpolating region and the UV given by $r={\cal R}_{D5/\overline{D5}}=\sqrt{3}a$ that acts as the effective boundary beyond which one does not generate a temperature-dependent UV-divergent counter term. For ${\cal R}_{D5/\overline{D5}}\gg r_0$, the trace anomaly will be generated from the  infinitesimal Weyl transformation (\ref{infinitesimal-Weyl-transf}) of the ${\cal R}_{D5/\overline{D5}}(\gg r_0)$-dependence in the counter term used for canceling the abovementioned ``divergent" ${\cal R}_{D5/\overline{D5}}^2\log {\cal R}_{D5/\overline{D5}}$ term (arising from$\int_{\rm IR} G_4\wedge *G_4$ ).

For the purpose of applying (\ref{trace-T}) to calculate $T^\mu_{\ \mu}$, we will be evaluating the same at the boundary $r={\cal R}_{D5/\overline{D5}}=\sqrt{3}a(\tilde{t})$ assuming that (\ref{trace-T}) is to be used with the same Weyl scalar $\sigma(x)$ (because $x\neq r$) at the boundary $r={\cal R}_{D5/\overline{D5}}$, as the flux integral in the UV $r>\sqrt{3}a(\tilde{t})$ is assumed to give a negligible contribution as compared to the IR. One can show that:
\begin{equation}
\label{S_FluxM}
 \int_{\rm IR} G_4\wedge *G_4 \sim \frac{ a^6 {g_s^{\rm IR}}\ ^{13/4} N^{\frac{3}{4}} M_{\rm IR}^4 {N_f^{\rm IR}}\ ^{\frac{5}{3}}  \log ({\cal R}_{D5/\overline{D5}})}{{\cal R}_{D5/\overline{D5}}^2\alpha _{\theta _1}^{11} \alpha _{\theta _2} \log
   ^{\frac{10}{3}}(N)}\int_{r\in{\rm IR}} d^4x \sqrt{g_{M_4}(x^{0,1,2,3})},
\end{equation}
and thus, using (\ref{trace-T}), implying the following conformal anomaly:
\begin{eqnarray}
\label{flux6}
& & \frac{{\cal T}^{\mu}_{\ \mu} }{T^4} (T<T_c)\sim \frac{g_s^{\rm IR}\ ^{\frac{5}{4}} M_{\rm IR}^3\alpha_{\theta_2} }{N^{\frac{9}{20}}N_f^{\rm IR}\ ^{\frac{1}{3}}\alpha_{\theta_1}^8
\left(\log N\right)^{\frac{16}{3}}\log(R_{\rm UV})} \left(\frac{a(\tilde{t})}{T}\right)^4 \left(\frac{a(\tilde{t})}{R_{\rm UV}}\right)^4\log \left(\sqrt{3}a(\tilde{t})\right),\nonumber\\
& & 
\end{eqnarray}
which using (\ref{GHY-cancel-FluxM_UV}) yields:
\begin{eqnarray}
\label{flux7}
& & \hskip -0.3in \frac{{\cal T}^{\mu}_{\ \mu} }{T^4} (T<T_c) \sim  \frac{g_s^{\rm IR}\ ^{\frac{5}{4}} M_{\rm IR}^3 }{g_s^{\rm UV}\ ^{\frac{3}{2}}N^{\frac{1}{4}}M_{\rm UV}\ ^{3}N_f^{\rm IR}\ ^{\frac{1}{3}}N_f^{\rm UV}\ ^{\frac{1}{3}}\left(\log N\right)^{\frac{1}{3}}\log(R_{\rm UV}) R_{\rm UV}^4} \left(\frac{a(\tilde{t})}{T}\right)^4\log \left(\sqrt{3}a(\tilde{t})\right).\nonumber\\
& & 
\end{eqnarray}

\begin{figure}[h]
 \begin{center}
%\begin{center}
 \includegraphics[scale=0.7]
 %[height= 21cm,width=+15cm]
 {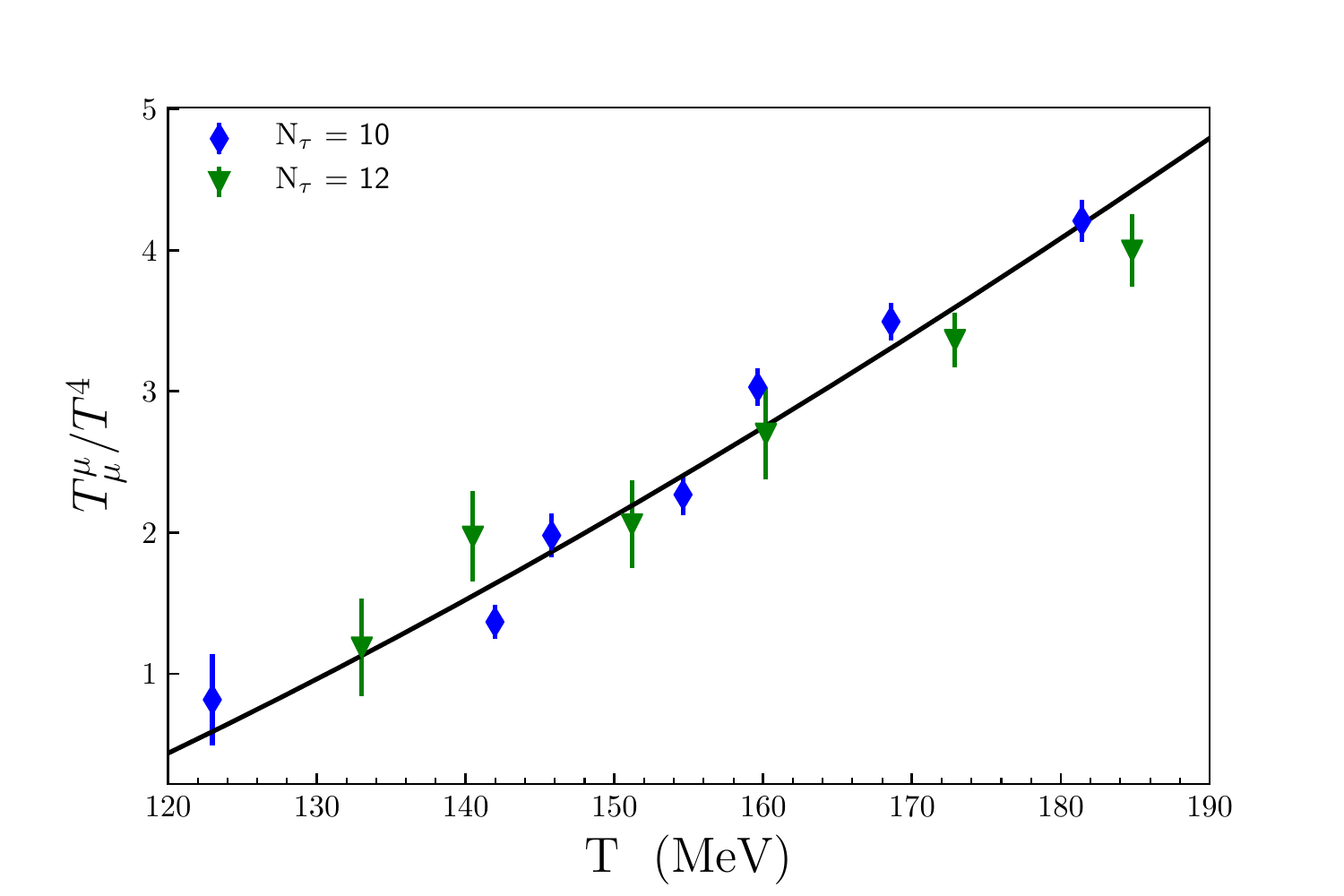}
 \end{center}
 \caption{The  trace anomaly, $\frac{T^{\mu}_{\ \mu} }{T^4}$, plotted as a function of temperature for the low temperature region. The points represent lattice QCD results (Fig. 2 of Reference \cite{Petreczky-et-al}). The different $N_\tau$ values refer to the number of lattice cells in the imaginary time direction. The full curve is the  chi-square minimization fit obtained with the approach described in {\bf 2.2}. The full curve represents the M theory result, Eqs. (\ref{flux6}) and (\ref{fits}).} 
  \label{Lattice_lowT}
\end{figure}

Further, assuming $\frac{a(\tilde{t})}{R_{\rm UV}}\sim \tilde{t}^{\frac{5}{4}}$ (inspired by the fractional temperature dependence of the Ouyang embedding parameter arising from holographic computation of electrical conductivity in \cite{EPJC-2}), and defining $T^\mu_{\ \mu} \equiv \log (R_{\rm UV})\left(T_c R_{\rm UV}\right)^4{\cal T}^{\mu}_{\ \mu}$ one obtains as the leading low-temperature contribution to the QCD trace anomaly:
\begin{eqnarray}
\label{conformal-anomaly-low-T}
& &\hskip -0.3in \frac{T^\mu_{\ \mu}}{T^4} \sim  \frac{g_s^{\rm IR}\ ^{\frac{5}{4}} M_{\rm IR}^3 }{g_s^{\rm UV}\ ^{\frac{3}{2}}N^{\frac{1}{4}}M_{\rm UV}\ ^{3}N_f^{\rm IR}\ ^{\frac{1}{3}}N_f^{\rm UV}\ ^{\frac{1}{3}}\left(\log N\right)^{\frac{1}{3}}}  \tilde{t} \log \left(\sqrt{3}\alpha_{a(\tilde{t})}\tilde{t}\right) \equiv \beta~ \tilde{t} \log \left(\sqrt{3}\alpha_{a(\tilde{t})}\tilde{t}\right).\nonumber\\
& & 
\end{eqnarray}
The aforementioned behavior of $a=a(\tilde{t})$ was to ensure in the simplest way that $T^\mu_{\ \mu}(T=0) = 0$ (as $T^\mu_{\ \mu}(T) =  G^{\mu\nu}G_{\mu\nu}  (T=0) -  G^{\mu\nu}G_{\mu\nu}  (T)$ \cite{trace_T_G^2(0)-G^2(T)}).
 We may now compare with results of lattice QCD calculations, in the ``low temperature'' sector, as reported in  \cite{Petreczky-et-al}. Combining the lattice data sets which span $0.120 < T < 0.19$ GeV ($N_\tau = 10, 12$), one obtains the agreement shown in Figure \ref{Lattice_lowT}. A good global chi-squared minimization fit is reached with: 
\begin{eqnarray}
\label{fits}
 \beta=7.04 \pm 0.5, \alpha_{a_0}=0.78 \pm 0.02.
\end{eqnarray}
%\begin{figure}[h]
 %\begin{center}
%\begin{center}
% \includegraphics[scale=0.8]
 %[height= 21cm,width=+15cm]
% {Trace-over-T^4-small-T-Nt=10.pdf}
 %\end{center}
 %\caption{$\frac{T^{\mu}_{\ \mu} }{T^4}$-vs-$T$,  for low temperatures and lattice temporal extent $N_\tau=10$ - the red plot is our M theory result (\ref{flux6}), (\ref{fits Ntau=10}) and the error-bar plot in blue is plotted using the data set from \cite{Petreczky-et-al}}
%\end{figure}

\subsection{${\cal O}(l_p^6 R^4)$ Terms ($r_h=0$)}

%The UV-divergent contribution in (\ref{flux2}) is canceled by an appropriate boundary counter term in ${\cal S}_{\rm ct}$ of (\ref{D11SUGRAS}) - see \cite{MQGP}.

Let us weigh in the higher derivative terms in the action (\ref{J+E_8+X_8}) up to ${\cal O}(R^4)$ (that are of ${\cal O}(l_p^6)$) in the $r_h=0$ limit. 

%In \cite{MQGP} it was shown that:
%\begin{eqnarray}
%\label{EH1}
%\sqrt{G^{\cal M}}R \sim \frac{r^3\cos\theta_2 \cot^2\theta_2\csc^4\theta_1f_2(\theta_2)}{g_s^{\frac{11}{4}} N^{\frac{3}{4}}},
%\end{eqnarray}
%$f_2(\theta_2)\sim\cot\theta_2$ near small values of $\theta_2$ \cite{NPB}. One hence notes that near small values of $\theta_{1,2}$ 
%($\theta_{1,2}\sim N^{-\frac{1}{5}}, N^{-\frac{3}{10}}$) 
%whereat the EH action receives the most dominant contribution, there is no large-$N$ finite UV-finite contribution  arising from $\int \sqrt{G^{\cal M}}R$. It is canceled by a suitable boundary counter terms in ${\cal S}_{\rm ct}$ (similar to \cite{MQGP}).
\begin{enumerate}
\item
First off, let us consider $J_0$. One can show that in the MQGP limit \cite{Yadav+Misra}:
\begin{eqnarray}
\label{J0-1}
J_0 \sim R^{\phi_2 r \theta_1 r}R_{\phi_1 r \theta_1 r}R_{\phi_2 r \phi_1 r}R^{\theta_1}_{\ \ r \theta_1 r} + \frac{1}{2} R^{\phi_2 r \theta_1 \theta_2}R_{r \psi \theta_1 r} R_{\phi_2}^{\ \ r \phi_1 r}
R^{\psi}_{\ \ r \phi_1 r},
\end{eqnarray}
and for $r_h=0$:
\begin{eqnarray}
\label{J0-2}
& & \int_{\rm UV}\sqrt{G^{\cal M}}J_0 \sim a^{4} \frac{ M^{\rm UV}\ ^2  N_{f\ {\rm UV}}^2\log N}{N^{\frac{13}{10}}}\left(\log a(\tilde{t})\right)^2,\nonumber\\
& & \int_{\rm IR}\sqrt{G^{\cal M}}J_0 \sim a^4 \frac{ M^{\rm IR}\ ^2  N_{f\ {\rm IR}}^2\log N \left(\log a(\tilde{t})\right)^2}{N^{\frac{13}{10}}}, 
\end{eqnarray}
implying $l_p^6\int_{\rm UV}\sqrt{G^{\cal M}}J_0 \ll l_p^6\int_{\rm IR}\sqrt{G^{\cal M}}J_0 \ll S_{\rm GHY}$.
We thus notice that this contribution, in the MQGP limit of \cite{MQGP},  is extremely suppressed relative to the one from the flux contribution (\ref{flux6}), and hence would be discarded.

\item
One can show that \cite{Yadav+Misra} (keeping track of only powers of $N$ and $r$-dependent terms), as a sample term:
\begin{eqnarray}
\label{E8-1}
& & E_8 \ni R^{tx^1}_{\ \ \ tx^1} R^{x^2x^3}_{\ \ \ \ x^2x^3} R^{r\theta_1}_{\ \ \ r\theta_1} 
R^{\psi x^{10}}_{\ \ \ \ \psi x^{10}} ,
\end{eqnarray}
and hence:
\begin{eqnarray}
\label{E8-2}
& & \int_{\rm IR}\sqrt{G^{\cal M}}E_8 \ni 10^{-6}\frac{ M_{\rm IR} 
 N_f^{\rm IR}  \left(\log N\right)^{\frac{14}{3}} 
{\cal R}_{D5/\overline{D5}}^4}{N^{\frac{31}{20}}\left(\log {\cal R}_{D5/\overline{D5}}\right)^{\frac{20}{3}}},\nonumber\\
& &  \int_{\rm UV}\sqrt{G^{\cal M}}E_8 \ni 10^{-6}\frac{ M_{\rm UV} 
 N_f^{\rm UV}  \left(\log N\right)^{\frac{14}{3}} 
R_{\rm UV}^4}{N^{\frac{31}{20}}\left(\log R_{\rm UV}\right)^{\frac{20}{3}}}.
\end{eqnarray}
Hence, from (\ref{trace-T}) one sees that the contribution from the $E_8$ term in the supergravity action would be extremely suppressed relative to the flux contribution and therefore will be discarded. 

\item  
Finally, let us look at the $t_8t_8G^2R^3$ term in (\ref{J+E_8+X_8}). One can show \cite{Yadav+Misra} that near $\theta_{1,2}\sim N^{-\frac{1}{5}}, N^{-\frac{3}{10}}$, the same is ${\cal O}\left(\frac{1}{N^{\frac{111}{20}}}\right)$, and hence is extremely suppressed as compared to the flux contribution, and will also be discarded.
\end{enumerate}

\section{Summary and Discussion}

In this paper, we showed that for  $T_c=150$ MeV, $N_f=3$, our M theory results for the variation of the QCD conformal anomaly with temperature can be made to be consistent with 
%NNLO HTL perturbation theory results at one loop at strong gauge coupling of \cite{QCD-Trace-Anomaly-lattice_etc} (which however uses $T_c=170 MeV, N_f=3$), and more 
recent lattice results of \cite{Petreczky-et-al}, for both,  high temperatures ($T>T_c$)  and low temperatures ($T<T_c$). 

The following points are noteworthy: 
\begin{enumerate}
\item
 As explained in the first reference in \cite{mesons_0E++-to-mesons-decays} in the context of obtaining meson spectroscopy consistent with hadronic phenomenology, even though obtaining the  type IIA mirror of \cite{metrics} and its M-theory uplift in \cite{MQGP} required a lot of work, but once obtained, we are able to obtain the trace/conformal anomaly from M theory which is very close to recent lattice results, already at ${\cal O}\left(M^0\right)$ (which is the non-conformal parameter) for $T>T_c$ \footnote{There is explicit dependence on $g_s, M, N_f, N$ of the conformal anomaly for $T<T_c$; when written in terms of the temperature, there is also an implicit dependence of the conformal anomaly on the aforementioned parameters via $r_h=r_H(T; g_s, M, N_f, N)$ for $T>T_c$.}. There are two major reasons why this happens.
\begin{itemize}
\item
 As noted explicitly in the first reference in \cite{mesons_0E++-to-mesons-decays}, the type IIA SYZ(Strominger-Yau-Zaslow) mirror of \cite{metrics} on account of the mixing of the type IIB metric and the NS-NS $B^{\rm IIB}$ under triple T duality, picks up sub-dominant terms in $N$ of
 ${\cal O}\left(\frac{1}{N^\kappa}\right), 0<\kappa<1$ which are also of ${\cal O}(M^0)$ in $B^{\rm IIA}$ (the non-conformality in type IIB in \cite{metrics} is because of $B^{\rm IIB}$ which is accompanied by $M$, the number of fractional $D3$ branes) which are therefore bigger than the ${\cal O}(\frac{g_sM^2}{N})$ contributions, that were missed, e.g.  in \cite{Dasgupta_et_al_Mesons} in the context of top-down meson spectroscopy.

\item
 The type IIB gravity dual  of \cite{metrics}  involved a resolved warped deformed conifold with a black hole and $\overline{D5}$, $D7$ branes and $\overline{D7}$ branes (plus fluxes). The type IIA mirror yields a non-K\"{a}hler warped resolved conifold with a black hole and $D6, \overline{D6}$ branes (plus fluxes).  Now, warped resolved conifolds are more easier to deal with computationally than  resolved warped deformed conifolds \footnote{One of us [AM] thanks K.Dasgupta for a short discussion on this point.}. It is the latter that are uplifted to M theory involving seven-folds of $G_2$ structure (See \cite{NPB}; the $D6$ branes get uplifted to KK monopoles).

\end{itemize}
\item
Strominger-Yau-Zaslow mirror construction is an entirely new technique  used  for studying QCD, holographically, from a top-down string/M-theory dual.

%\item
%The type IIB non-conformality markers, $M$ - the number of fractional (anti-)$D3$-branes - and $N_f$ - the number of flavor (anti-)$D7$-branes, appear in our holographic computations of the trace anomaly via the horizon radius in the high-temperature limit ($T>T_c$) in subsection {\bf 2.1}, and as an explicit pre-factor in the low-temperature limit ($T<T_c$) in subsection {\bf 2.2}.

\item
The main Physics lesson that one learns in this work is the following. It turns out that for high temperatures ($T>T_c$) corresponding to a black-hole in the M theory dual, it is the counter term used to cancel the UV-divergent contribution of the GHY surface term (via reparametrization of the UV boundary) that contributes to the temperature dependence of the trace anomaly and guarantees vanishing of $\frac{T^{\mu}_{\ \mu} }{T^4}$ at asymptotically large temperatures. On the other hand, for low temperatures ($T<T_c$) corresponding to a thermal background (no black hole), it is the counter term used to cancel the divergent contribution arising from the flux term (at the boundary common to the IR-UV interpolating region and the UV [assuming the $D5-\overline{D5}$ separation to be much greater than the IR cut-off of the thermal background]) that contributes to the temperature dependence of the conformal anomaly, and guarantees increase of $\frac{T^{\mu}_{\ \mu} }{T^4}$ up to around $T_c$ with increase in temperature - just like lattice calculations \cite{Petreczky-et-al}.

\item
One should make note of the fact that was also stated  earlier in Section {\bf 1}, the type IIB string theory dual of thermal QCD as constructed in \cite{metrics}, unlike its earlier type IIA cousin - the Sakai-Sugimoto that catered only to the IR - is UV complete. So is hence the type IIA SYZ mirror constructed in \cite{MQGP}. Further, in the MQGP limit, the contributions of the higher order derivative corrections to the trace anomaly, will be severely large-$N$ suppressed. These two together ensure that the results of this paper on the temperature variation of the trace anomaly consistent with very recent lattice results,  obtained from a top-down approach unique to our work,  can be completely trusted.
\end{enumerate}

We have obtained the trace of the energy-momentum tensor in a top-down non-conformal M theory holographic dual which has a temperature behaviour consistent with that of QCD. In the approach outlined in this paper, the low and high temperature regions correspond to two different limits of the same theory, as is the case in the lattice QCD counterpart. 
%A strong word of caution is in order, however. It is well known that the phase diagram germane to the string theoretical large-$N_c$ approaches are ${\it not}$ to be rigorously interpreted as identical to the QCD phase diagram. Differences, both subtle and important, exist \cite{He:2013qq} and need to be carefully analyzed.

{\it To conclude}: There have been several papers on a holographic computation of the trace anomaly, but {\it all} bottom-up (e.g. \cite{Nitti_et_al}, \cite{trace-anomaly-bottum-up-2}, etc.). To the best of our knowledge, \cite{MQGP} is the only (top-down) holographic M-theory dual (of thermal QCD) that is able to yield (as shown in this paper):
\begin{itemize}
\item
(after a tuning of the (small) Ouyang embedding parameter and radius of a blown-up $S^2$ when expressed in terms of the horizon radius) a deconfinement temperature $T_c$ from a Hawking-Page phase transition at vanishing baryon chemical potential consistent with the very recent lattice QCD results in the heavy quark 
limit

\item
a conformal anomaly variation with temperature compatible with the very recent lattice results at  high ($T>T_c$) {\it and} low ($T<T_c$) temperatures - the latter missing, e.g., in the bottom-up \cite{Nitti_et_al} (apart from the fact that the authors compared with much older lattice results)

\end{itemize}
as well as (shown in earlier papers in the past few years):
\begin{itemize}
\item
Condensed Matter Physics: inclusive of  the non-conformal corrections, to obtain:
\begin{enumerate}
\item
 a lattice-compatible shear-viscosity-to-entropy-density ratio (first reference in \cite{EPJC-2}) 

\item
temperature variation of a variety of transport coefficients including the bulk-viscosity-to-shear-viscosity ratio,  diffusion coefficient, speed of sound (the last reference in \cite{EPJC-2}), electrical and thermal conductivity and the Wiedemann-Franz law (first reference in \cite{EPJC-2});
\end{enumerate}

\item
Particle Phenomenology:  obtaining:
\begin{enumerate}
\item
 lattice compatible glueball spectroscopy \cite{Sil+Yadav+Misra-glueball} 

\item
meson spectroscopy (first reference of \cite{mesons_0E++-to-mesons-decays}) 

\item
glueball-to-meson decay widths (second reference of \cite{mesons_0E++-to-mesons-decays})
\end{enumerate}

\item
Mathematics: using the beautiful concept of (SYZ) Mirror Symmetry in algebraic geometry, and the machinery of G-structures to provide, for the first time, an $SU(3)$-structure (for type IIB (second reference of \cite{EPJC-2})/IIA \cite{NPB}   holographic dual) and $G_2$-structure  \cite{NPB}  torsion classes of the six- and seven-folds relevant to top-down holographic duals of thermal QCD.

\end{itemize}
 
The results of this note do demonstrate the potential of the methods outlined here to treat strongly-coupled problems analytically, whether conformal symmetry is manifest or not. Other applications to hadronic physics will include studies of the possible critical point in the hadronic phase diagram  and of the high-density/low temperature color superconducting phase.

\section*{Acknowledgement}

This work was supported in part by the Natural Sciences and Engineering Research Council of Canada. 
We thank  K. Dasgupta for many useful discussions, and 
%M.Strickland for sharing the data set for the HTL 1-loop plot in \cite{QCD-Trace-Anomaly-lattice_etc}, F.~Karsch and S.~Mukherjee for bringing \cite{lattice-low+high-T-trace-over-T^4} to our attention and 
P.~Petreczky for pointing out the location of the lattice data shown in Ref. \cite{Petreczky-et-al}, AM would like to thank the McGill high energy theory group (K. Dasgupta in particular) for the wonderful hospitality during his visits to the same, at various stages of this work, and M.~Dhuria for help with some plots. AM was partially supported by a grant from the Council of Scientific and Industrial Research,  Government of India, grant number CSR-1477-PHY.

\appendix
\section{The Baryon Chemical Potential $\mu_C$ and the DBI Action on the Flavor $D7$-Branes \label{seceqbb}}
\setcounter{equation}{0}\seceqaa

In this appendix we discuss the evaluation of the baryon chemical potential $\mu_C$ and the DBI action on the world volume of the flavor $D7$-branes. This is relevant to the discussion of showing lattice compatibility of our supergravity computation of $T_c$ in section {\bf 3}.

It was shown in \cite{NPB} (by turning on a world-volume flux $F_{rt}=\partial_r A_t(r)$) that the baryon chemical potential $\mu_C$ is given as under (with the simplifying assumption that using the Ouyang embedding (\ref{Ouyang-definition}): $e^{-\phi} = \frac{1}{g_s} - \frac{N_f^{\rm IR}}{2\pi}\log|\mu_{\rm Ouyang}|$ for $r\in[r_h,\sqrt{3}a]$ and $e^{-\phi} = \frac{1}{g_s} - \frac{N_f^{\rm UV}}{2\pi}\log|\mu_{\rm Ouyang}|$ for $r\in(\sqrt{3}a,R_{UV}]$):
   \begin{eqnarray}
   \label{muC-1}
   & &  \mu_C = \int_{r_h}^\infty F_{rt}dr = 
  - \int_{r_h}^\infty dr\frac{C}{\sqrt{C^2+r^{9/2} \left(\frac{{N_f(r)} \log|\mu_{\rm Ouyang}|}{2 \pi
   }-\frac{1}{{g_s}}\right)^2}}\nonumber\\
   & &  \sim -\int_{r_h}^{R_{\rm UV}}dr\frac{d}{dr} \left[r \ _2F_1\left(\frac{2}{9}, \frac{1}{2}; \frac{11}{9};-\frac{r^{9/2} \left(\frac{{Nf(r)} \log
   |\mu_{\rm Ouyang}|}{2 \pi } - \frac{1}{{g_s}}\right)^2}{C^2}\right)\right],
\end{eqnarray}
which for $|\mu_{\rm Ouyang}|\gg1$ yields:   
\begin{eqnarray}
\label{muC-2}
& &    -\frac{2^{4/9} C^{4/9} \Gamma \left(\frac{5}{18}\right) \Gamma \left(\frac{11}{9}\right)\left(\frac{1}{{N_f^{\rm UV}}^{4/9}}-\frac{1}{{N_f^{\rm IR}}^{4/9}}\right)}{\left(\log|\mu_{\rm Ouyang}|\right)^{\frac{4}{9}}\sqrt[18]{\pi }}\ 
\end{eqnarray}
if $\mu_{\rm Ouyang}$ does not depend on $r_h$,
and
\begin{eqnarray}
\label{muC-3}
& &    -\frac{2^{4/9} C^{4/9} \Gamma \left(\frac{5}{18}\right) \Gamma
   \left(\frac{11}{9}\right)}{\sqrt[18]{\pi } {N_f^{\rm UV}}^{4/9} \log ^{\frac{4}{9}}(|\mu_{\rm Ouyang}|}-\frac{{N_f}^2 {r_h}^{11/2} \log ^2(\mu )}{44 \pi ^2 C^2}+{r_h}
   \end{eqnarray}
if $|\mu_{\rm Ouyang}| \sim {r_h^{-\alpha}}, \alpha>0 $ \cite{EPJC-2}.

Similarly, when $|\mu_{\rm Ouyang}|\ll1$, one obtains:
 \begin{eqnarray}
 \label{muC-4}
\mu_C =    -\frac{2^{4/9} C^{4/9} \Gamma \left(\frac{5}{18}\right) \Gamma \left(\frac{11}{9}\right)\left(\frac{1}{{N_f^{\rm UV}}^{4/9}}-\frac{1}{{N_f^{\rm IR}}^{4/9}}\right)}{\left(-\log|\mu_{\rm Ouyang}|\right)^{\frac{4}{9}}\sqrt[18]{\pi }}
\end{eqnarray}
if $\mu_{\rm Ouyang}$ does not depend on $r_h$, and:
\begin{eqnarray}
\label{muC-5}
& & \mu_C =   -\frac{2^{4/9} C^{4/9} \Gamma \left(\frac{5}{18}\right) \Gamma
   \left(\frac{11}{9}\right)}{\sqrt[18]{\pi } {N_f^{\rm UV}}^{4/9} \left(-\log|\mu_{\rm Ouyang}|\right)^{\frac{4}{9}}}+\frac{{N_f^{\rm IR}}\ ^2 {r_h}^{11/2} \log ^2|\mu_{\rm Ouyang}|}{44 \pi ^2 C^2}-{r_h}
\end{eqnarray}   
if $|\mu_{\rm Ouyang}| \sim {r_h^{-\alpha}}, \alpha>0$ \cite{EPJC-2}.

Further, one notes that the $D7$-brane DBI action will be given by:
\begin{eqnarray}
\label{SDBI-i}
& &  S_{\rm DBI}^{\rm UV-finite} \sim T_{D7} \int_{r={\cal R}_{D5/\overline{D5}}}^{R_{UV}} \sqrt{\mu } r^{9/4} \left(\frac{1}{{g_s}}-\frac{{N_f(r)} \log |\mu_{\rm Ouyang}|}{2 \pi
   }\right)^2 \sqrt{\frac{r^{9/2}}{C^2+r^{9/2} \left(\frac{1}{{g_s}}-\frac{{N_f(r)}
   \log |\mu_{\rm Ouyang}|}{2 \pi }\right)^2}}\nonumber\\
   & &  \sim  T_{D7} \int_{\sqrt{3}a}^{R_{UV}}dr \frac{d}{dr}\biggl[\frac{2 \sqrt{\mu } r \sqrt{4 \pi ^2 \left(C^2 {g_s}^2+r^{9/2}\right)+{g_s}^2
   {N_f(r)}^2 r^{9/2} \log ^2|\mu_{\rm Ouyang}|-4 \pi  {g_s} {N_f(r)} r^{9/2} \log|\mu_{\rm Ouyang}|}
   }{13 \pi 
   {g_s}^2}\nonumber\\
   & & \times \left({g_s}-2 \pi  C {g_s}\ _2F_1 \left(\frac{2}{9}, \frac{1}{2};\frac{11}{9};-\frac{r^{9/2} ({g_s}
   {N_f(r)} \log |\mu_{\rm Ouyang}|-2 \pi )^2}{4 C^2 {g_s}^2 \pi ^2}\right)\right)\biggr].
   \end{eqnarray}
   The ``UV-finite part" of (\ref{SDBI-i}), i.e., the action that remains finite in the large $R_{UV}$-limit will be given by: 
  \begin{eqnarray}
  \label{SDBI-ii}
   & & S^{\rm Large\ R_{UV}-finite}_{DBI} \sim
  -\frac{2^{22/9} \sqrt{\mu } \Gamma \left(\frac{5}{18}\right) \Gamma
   \left(\frac{11}{9}\right) \left(\frac{C}{{N_f^{\rm UV}}}\right)^{4/9}}{13 \sqrt[18]{\pi
   } \log|\mu_{\rm Ouyang}| ^{\frac{4}{9}}} -\frac{3^{11/4} b^{11/2} {N_f}^2 {r_h}^{\frac{(11 - \alpha)}{2}} \log
   ^2({r_h})}{22 \pi ^2 C},\nonumber\\
   & & 
   \end{eqnarray}
for    $ |\mu_{\rm Ouyang}|\gg1$ (assuming $|\mu_{\rm Ouyang}| \sim r_h^{-\alpha}, 0<\alpha<11$) and:
\begin{eqnarray}
\label{SDBI-iii}
 & &  S^{\rm Large\ R_{UV}-finite}_{DBI} \sim -\frac{2^{22/9} C^{13/9} \sqrt{\mu } \Gamma \left(\frac{5}{18}\right) \Gamma
   \left(\frac{11}{9}\right)}{13 \sqrt[18]{\pi } {N_f^{\rm UV}}^{4/9} |\mu_{\rm Ouyang}|^{\frac{4}{9}} } - \frac{243 \times \sqrt[4]{3} a^{11/2} \sqrt{\frac{1}{C^2}} \sqrt{\mu } {N_f}^2 \log ^2(\mu
   )}{22 \pi ^2},\nonumber\\
   & & 
\end{eqnarray}   
for   $|\mu_{\rm Ouyang}|\ll1$. 

\section{Definitions of symbols in the $D=11$ supergravity action  (\ref{J+E_8+X_8}) \label{seceqaa}}
\setcounter{equation}{0} \seceqbb

In  (\ref{J+E_8+X_8}),  $(J_0,E_8,X_8)$ are quartic polynomials in the curvature tensor in 11-dimensional space and defined as:
\begin{eqnarray}
\label{J+E_8+X_8}
& & J_0 = 3.2^8 \Bigl(R^{MIJN}R_{PIJQ} R_{M}^{\ RSP}R^{Q}_{\ RSN}+\frac{1}{2}R^{MNIJ}R_{PQIJ} R_{M}^{\ RSP}R^{Q}_{\ RSN}\Bigr),\nonumber\\
& & E_8= \epsilon^{PQR M_1 N_1...M_4 N_4} \epsilon_{PQR M^{\prime}_1 N^{\prime}_1...M^{\prime}_4 N^{\prime}_4}R^{M^{\prime}_1 N^{\prime}_1}_{\ \ \ \ \ \ M_1 N_1}...R^{M^{\prime}_4 N^{\prime}_4}_{\ \ \ \ \ \ M_4 N_4},\nonumber\\
& &  X_8 = \frac{1}{192 \cdot (2 \pi^2)^4}\Bigl[tr(R^4)-\frac{1}{4}(tr{R^2})^2\Bigr],
 \end{eqnarray}
where $M, N, P,...$ are $D=11$ indices; the $t_8$-symbol is defined as (see  \cite{Grimm_et_all-O(R^4)}):
\begin{eqnarray}
t_8^{N_1\dots N_8}   &=& \frac{1}{16} \big( -  2 \left(   G^{ N_1 N_3  }G^{  N_2  N_4  }G^{ N_5   N_7  }G^{ N_6 N_8  }
 + G^{ N_1 N_5  }G^{ N_2 N_6  }G^{ N_3   N_7  }G^{  N_4   N_8   }
 +  G^{ N_1 N_7  }G^{ N_2 N_8  }G^{ N_3   N_5  }G^{  N_4 N_6   }  \right) \nonumber \\
 & &  +
 8 \left(  G^{  N_2     N_3   }G^{ N_4    N_5  }G^{ N_6    N_7  }G^{ N_8   N_1   }
  +G^{  N_2     N_5   }G^{ N_6    N_3  }G^{ N_4    N_7  }G^{ N_8   N_1   }
  +   G^{  N_2     N_5   }G^{ N_6    N_7  }G^{ N_8    N_3  }G^{ N_4  N_1   }
\right) \nonumber \\
& &  - (N_1 \leftrightarrow  N_2) -( N_3 \leftrightarrow  N_4) - (N_5 \leftrightarrow  N_6) - (N_7 \leftrightarrow  N_8) \big) \,.
\end{eqnarray}
Also, from \cite{G^2R^3}:
\begin{eqnarray}
\label{G^2R^3}
t_8t_8G^2R^3 = t_8^{M_1...M_8}t^8_{N_1....N_8}G_{M_1}\ ^{N_1 PQ}G_{M_2}\ ^{N_2}_{\ \ PQ}R_{M_3M_4}^{\ \ \ \ N_3N_4}R_{M_5M_6}^{\ \ \ \ N_5N_6}R_{M_7M_8}^{\ \ \ \ N_7N_8}.
\end{eqnarray}

\end{document}